\documentclass[preprint,showpacs,preprintnumbers,amsmath,amssymb]{revtex4}

\usepackage{graphicx,color}
\usepackage{dcolumn}
\usepackage{bm}

\begin{document}

\title{Study of cosmic ray interaction model based on atmospheric muons for the neutrino flux calculation}

\author{T.~Sanuki}
\email[]{sanuki@icepp.s.u-tokyo.ac.jp}
\affiliation{International Center for Elementary Particle Physics, the University of Tokyo, 7-3-1 Hongo, Bunkyo-ku, Tokyo 113-0033, Japan}
\author{M.~Honda}
\email[]{mhonda@icrr.u-tokyo.ac.jp}
\homepage[]{http://icrr.u-tokyo.ac.jp/~mhonda}
\author{T.~Kajita}
\email[]{kajita@icrr.u-tokyo.ac.jp}
\affiliation{Institute for Cosmic Ray Research, the University of Tokyo, 5-1-5 Kashiwa-no-ha, Kashiwa, Chiba 277-8582, Japan}
\author{K.~Kasahara}
\email[]{kasahara@icrc.u-tokyo.ac.jp}
\affiliation{Shibaura Institute of Technology, 307 Fukasaku, Minuma-ku, Saitama 337-8570, Japan}
\author{S.~Midorikawa}
\email[]{midori@aomori-u.ac.jp}
\affiliation{Faculty of Software and Information Technology, Aomori University,  2-3-1 Kobata, Aomori, Aomori 030-0943, Japan.}

\date{\today}

\begin{abstract}
We have studied the hadronic interaction for the calculation 
of the atmospheric neutrino flux 
by summarizing the accurately measured atmospheric muon flux data
and comparing with simulations.
We find the atmospheric muon and neutrino fluxes respond to errors 
in the $\pi$-production of the hadronic interaction similarly, 
and compare the atmospheric muon flux calculated using the 
HKKM04~\cite{HKKM2004} code 
with experimental measurements.
The $\mu^++\mu^-$ data show good agreement in the 1$\sim$30~GeV/c range,
but a large disagreement above 30~GeV/c.
The $\mu^+/\mu^-$ ratio shows sizable differences at lower 
and higher momenta for opposite directions.
As the disagreements are considered to be due to assumptions in the hadronic 
interaction model,
we try to improve it phenomenologically based on the quark parton model.
The improved interaction model reproduces the observed muon flux data well.
The calculation of the atmospheric neutrino flux will be 
reported in the following paper~\cite{hkkms2006}.
\end{abstract}

\maketitle

\section{\label{sec:introduction} Introduction}

Evidence of neutrino oscillations was found in the atmospheric 
neutrino data observed with Super-Kamiokande \cite{Fukuda:1998mi}.
Atmospheric neutrinos are still a powerful tool to study  
neutrino oscillations, 
and the overall uncertainties in the observed data are
becoming smaller \cite{Ashie:2005ik}.
It is highly desirable to predict the absolute flux value and
ratios among different kind of neutrinos precisely, and
to understand their ``systematic'' uncertainties.
Note, atmospheric neutrino experiments cover a wide 
$L/E(=[neutrino~flight~length]/[neutrino~energy])$ range, 
over four orders of magnitude \cite{Ashie:2004mr},
which is much wider than accelerator neutrino beam 
experiments such as K2K \cite{Aliu:2004sq}.
Atmospheric neutrino experiments are 
complementary to accelerator neutrino experiments, 
which enable a narrow parameter region to be accurately surveyed.

In order to calculate the atmospheric neutrino intensities precisely, 
we need detailed information about 
(i) the primary cosmic-ray spectra at the top of the atmosphere,
(ii) the hadronic interactions between cosmic rays and atmospheric nuclei, 
(iii) the propagation of cosmic-ray particles inside the atmosphere, and
(iv) the decay of the secondary particles.

For (i) the primary cosmic spectra,
the uncertainties have been greatly reduced with new measurements of 
primary cosmic rays \cite{AMS1p, AMS1He, BESSpHe, caprice98-p}.
Among these experiments, the spectra of cosmic ray protons reported by AMS and BESS show 
a very good agreement up to around 100~GeV, 
although they were carried out in very different experimental conditions.
AMS flew on-board the space shuttle orbiting at the altitudes 
between 320~km and 390~km.
On the other hand, 
BESS was a balloon-borne experiment carried out at the atmospheric depths 
of about 5~g/cm$^2$ ($\sim$~37~km a.s.l.).
Then BESS-TeV \cite{BESSTeVpHemu}, 
the upgraded BESS experiment which extended the energy region up to 540~GeV,
confirmed the results of AMS and BESS.
The results of BESS-TeV agree with that of BESS to within 3\%.

We note the observed proton spectrum by CAPRICE is obviously lower than that
of AMS or BESS. However, the event number acquired by AMS and BESS is far 
larger than that of CAPRICE. Although it is difficult to combine the results of 
the different experiments, a combined analysis using AMS, BESS and CAPRICE
data would gives a very close result to that using AMS and BESS 
only~\cite{Gaisser-hamburg}.
The difference of Helium spectra observed by AMS and BESS are also sizable.
However, 
as the proton is the dominant component in the cosmic rays,
the difference in terms of the nucleon flux is less than $\sim$4\%
below 100 GeV.
The nucleon flux is important in the calculation of atmospheric
muon and neutrino fluxes. 
We consider that the cosmic ray is well understood below 100~GeV,
which is important for the calculation of atmospheric muons and 
neutrinos below 10 GeV.

For the study of (ii) the hadronic interactions, 
the accelerator experiment is the most direct method.
However, the data available now do not cover all the phase space 
necessary for the calculation of atmospheric neutrino flux.
We study the hadronic interactions using 
the atmospheric muon flux data in this paper.
As the energy of $\pi$ or $K$ mostly goes to muons at their decay, 
the muons are considered to carry essential information 
of $\pi$ and $K$ production in the hadronic interactions.
There have been a lot of measurements of atmospheric muon flux
at ground level as compiled in Ref.~\cite{Grieder-book} and at the 
balloon altitudes~\cite{mass-mu,mass2-mu,caprice94-mu,caprice98-mu,heat-mu}.
Among them, we select the series of precise measurements of 
atmospheric muons by BESS at various altitudes;
sea level \cite{BESSTeVpHemu}, mountain altitude \cite{BESSnorimu} and
balloon altitude \cite{Abe:2003cd}, with 
sufficiently small systematic and statistical errors.
In all these measurements, they used essentially the same apparatus 
as for the primary cosmic-ray measurements \cite{BESSpHe,BESSTeVpHemu},
and systematic errors were well controlled.
There are other precision measurements of the atmospheric muon flux 
useful for the study in this paper,
such as the L3+C experiment~\cite{l3+c}.

Our study might be compared with the direct calculation of
atmospheric neutrino flux from the atmospheric muon 
flux~\cite{Zatsepin-kuzmin:1961,Hayakawa-book,tam-yan}, in which
the $\pi$'s are assumed as the dominant source of the atmospheric 
neutrinos and muons. The calculation was reviewed
in Ref.~\cite{Perkins:1994pm}.
The differences of our study from those works are in the use of a 
well-constructed model of primary cosmic ray spectra, 
and a full Monte Carlo simulation code for atmospheric muon and 
neutrino fluxes calculation.
Then we study the hadronic interactions comparing the calculated
and observed atmospheric muon fluxes.
Both the model of primary cosmic ray spectra and the simulation code are 
the same as those used in HKKM04 calculation~\cite{HKKM2004}.
The primary flux model is very close to that constructed in 
Refs.~\cite{Gaisser-hamburg,Gaisser-Honda} based on the AMS and BESS data.
We use the primary flux model with a modification of the spectrum index
of cosmic ray protons from -2.74 to -2.71 above 100~GeV, 
according to the emulsion chamber experiments at higher 
energies~\cite{JACEE-1998, RUNJOB-2005}.
The simulation code treat the (iii) propagation of cosmic-ray particles 
inside the atmosphere and (iv) decay of the secondary particles sufficiently 
accurately. 

Calculating the atmospheric neutrino flux from the primary cosmic ray flux, 
the atmospheric muon flux was also used to calibrate the calculation
in Ref.~\cite{Fiorentini:2001wa}.
However, the atmospheric density profile is crucial in this study.

First, we review the precision measurements of muon flux by BESS 
and other instruments in Sec.~\ref{sec:obs-mu}.
Next, we study what kind of information can be deduced from the 
comparison of the calculation of atmospheric muon flux and observed data 
for the calculation of atmospheric neutrino flux 
in Sec.~\ref{sec:mn-relation}.
We also study the effect of the atmospheric density profile
on the muon flux in Sec.~\ref{sec:atmosphere} before the comparison.
Note, the seasonal change of the air density profile 
causes $\pm$~5~\% variations of muon flux at $\sim$~1~GeV/c at sea level.
And even larger variation is expected to result from changes in local 
meteorological conditions.
In the same section we also discuss the effect of the uncertainty 
of the interaction cross-section between cosmic rays and air nuclei.

We calculate the muon fluxes in the HKKM04 scheme with the observed 
atmospheric density profile,
and compared them with the precisely measured muon flux data,
in Sec.~\ref{sec:comparison}.
Note, the existence of precise atmospheric density profile data during
the observation period is another important reason that we use the muon 
flux data from the BESS measurements.
We find $\mu^+ + \mu^-$ shows reasonable agreement 
in the 1$\sim$30~GeV/c range between the calculations and observations,
but that the agreement worsens above 30~GeV/c.
In addition, the $\mu^+ /\mu^-$ ratio shows a sizable difference. 
The difference is considered to be due to errors in the hadronic 
interaction model used in HKKM04 (DPMJET-III~\cite{Roesler:2000he}).

We try to improve the hadronic interaction model in 
Sec.~\ref{sec:modification}, and compared with the data
from recent accelerator data~\cite{na49-data}.
With a phenomenological consideration based on quark parton model,
$K$ productions are also modified in this ``improvement''.
As the result of the modification, the observed muon fluxes are 
reproduced with good accuracy (Sec.~\ref{sec:modified-calculation}).

Note, the available precision muon flux data are essentially those for 
the vertical directions.
If we have the accurately measured horizontal muon flux data,
we can test the simulation code by the comparison of the calculated 
and observed muon fluxes for the horizontal directions.
This comparison would be a good support for our procedure.
However, the muon flux data for horizontal direction are poorer than 
those for vertical directions. We just show the comparison of the 
calculation and available muon flux data for horizontal 
directions~\cite{Allkofer:1985ey,Matsuno:1984kq} in 
Sec.~\ref{sec:modified-calculation}.
The calculation of atmospheric neutrino flux with the 
modified interaction model will be reported 
in the following paper~\cite{hkkms2006}.

\section{\label{sec:obs-mu}Precision measurements of atmospheric muon}
The BESS group performed a series of atmospheric muon observations at 
various levels and sites; 
balloon altitude, mountain altitude and at sea level.

At balloon altitudes, where the atmospheric depths 
(5$\sim$25~g/cm$^2$) are much smaller than the interaction 
mean free path of protons ($\sim$~100~g/cm$^2$),
the muon flux measurement is considered as an inclusive experiment with 
the primary cosmic ray beam and the air nucleon target.
We can expect rich information about hadronic interactions from this region.
However, 
only a small number of experiments had been performed,
and their data are poor in statistics,
because the muon flux itself is small at balloon altitudes
and the observation time is limited.
On January 24th, 2001, the BESS group carried out 
a muon flux observation at balloon altitudes with exceptionally
good statistics at Fort Sumner, NM, USA~\cite{Abe:2003cd}.
After reaching an altitude with a residual atmospheric depth of 5~g/cm$^2$,
the balloon slowly descended to 28~g/cm$^2$ in 12.4 hours. 
A large number of primary and secondary cosmic rays were recorded
during the descending period.
The positive muon spectrum was obtained for 0.50--2.55~GeV/c and 
negative muon spectra for 0.50--9.76~GeV/c.
As discussed in Ref.~\cite{Abe:2003ah},
we selected DPMJET-III as the interaction model for HKKM04
with these muon data, since it
reproduced the observed atmospheric muon spectra
better than the interaction models of
Fritiof 1.6~\cite{Nilsson-Almqvist:1986rx}, 
Fritiof 7.02~\cite{Pi:1992ug}, 
and FLUKA'97~\cite{Fasso:2000hd}.

The BESS group has also measured the atmospheric muon flux for 
vertical directions at ground level;
at Mt.\ Norikura, Japan (742~g/cm$^2$) in September, 1999~\cite{BESSnorimu} 
and at Tsukuba, Japan (1032~g/cm$^2$) in October, 2002~\cite{BESSTeVpHemu}.
The momentum ranges covered are 0.58--106~GeV/c at Norikura
and 0.58--404~GeV/c at Tsukuba.
In both experiments, the observation times were long enough
that the systematic errors dominate the statistical errors.
The overall errors were 
3~\% at 1~GeV/c, 3~\% at 10~GeV/c, and 9~\% at 100~GeV/c for the Norikura experiment, 
2~\% at 1~GeV/c, 3~\% at 10~GeV/c, and 5~\% at 100 GeV/c for the Tsukuba experiment.
Note that
the Tsukuba experiment was carried out with the BESS-TeV detector,
which could not distinguish electrons and positrons from muons~\cite{BESSTeVpHemu}.
On the other hand, the BESS detector used for the Norikura experiment 
was equipped with an electromagnetic shower counter to distinguish electrons and 
positrons from muons~\cite{BESSnorimu}.
As an important aspect of the muon measurement by BESS at ground level,
the precise atmospheric density profile data are available from the Japan 
Meteorological Agency~\cite{koso-kishou}.

The L3+C detector has measured the atmospheric muon flux accurately from 
20~GeV/c to 3~TeV/c and at zenith angles from 0$^\circ$ to 58$^\circ$ 
at CERN~\cite{l3+c}.
The L3+C detector was originally constructed for the LEP experiment as 
the L3 muon spectrometer,
and the efficiency and the absolute momentum scale were calibrated with 
the muon pairs from $Z$ decays.
The overall error is read from Ref.~\cite{l3+c} to be 
4.5~\% at 20~GeV/c, less than 3~\% in 60 -- 500~GeV/c, and 10~\% at 1.5~TeV/c.
However, the L3+C detector is situated below a molasses overburden 
of 30~m (6854~g/cm$^2$), 
which could be a source of unknown systematic error at lower momenta. 
In the following study, 
we used the L3+C data for vertical directions ($\cos\theta_{zenith}~>~0.9$),
and in the momentum range of 60~GeV/c -- 3~TeV/c.

For the horizontal directions, there are muon flux data from 
the MUTRON~\cite{Matsuno:1984kq} 
and DEIS~\cite{Allkofer:1985ey} experiments at sea level.
MUTRON observed the muon flux from 100~GeV/c to 20~TeV/c in momentum 
and from 86$^\circ$ to 90$^\circ$ (88.9$^\circ$ on average) 
in zenith angle, 
and DEIS from 10~GeV/c to 10~TeV/c in momentum and from 
78$^\circ$ to 90$^\circ$ in zenith angle.
However, it is difficult to read the systematic errors 
from their reports~\cite{Matsuno:1984kq,Allkofer:1985ey}.
These data are potentially useful to study the validity 
of the simulation code at higher energies, rather than to
study the hadronic interaction model.

\section{\label{sec:mn-relation}Response of Atmospheric Muon and Neutrino fluxes to errors in hadronic interactions}

Before a study of the hadronic interactions,
we look at the response of atmospheric muon and neutrino fluxes 
to errors in the hadronic interaction model.
If their responses are the same,
we can study the hadronic interactions relevant to 
the atmospheric neutrino flux with the atmospheric muon flux data.
Here,
we derive some analytical expressions for the calculation of atmospheric
muon and neutrino fluxes, 
but we actually calculate them in the Monte Carlo simulation,
then interpret the results with the analytical expressions. 

We use the HKKM04 calculation code for the Monte Carlo calculation of 
atmospheric neutrino flux~\cite{HKKM2004}.
In the HKKM04 scheme,  
the primary cosmic ray flux model based on the AMS and BESS 
observations~\cite{Gaisser-hamburg, Gaisser-Honda} was used 
with a modification of the spectrum index for proton cosmic rays 
from $-$2.74 to $-$2.71 above 100~GeV, 
so that the extension goes through the center of the
emulsion experiment data~\cite{JACEE-1998, RUNJOB-2005} at higher energies.
DPMJET-III~\cite{Abe:2003ah} was selected for the hadronic interaction model, 
and the US-Standard '76~\cite{us-standard} atmospheric density profile was used.
Note, however,
the discussion in this section is not sensitive to the details
of the Monte Carlo simulation scheme.

To cut out the hadronic interactions,
we write the  atmospheric lepton ($\mu$, $\nu_\mu$, or $\nu_e$) fluxes as,
\begin{equation}
\label{eq:meson-lepton}
\phi_l(p_l) = 
\sum_{m=\pi^\pm,K^\pm, K^0_L, K^0_S}\int_h T_m^l(p_m, p_l, h)\cdot Y^m(p_m, h)~ dp_m dh~,
\end{equation}
where $l$ stands for the kind of lepton, 
$m$ stands for the kind of meson ($\pi^\pm, K^\pm, ...$),
$T_m^l(p_m, p_l, h)$ is the probability 
with which the $m$-meson produced with momentum $p_m$ and at altitude $h$ 
creates an $l$-lepton with momentum $p_l$ at ground level,
and $Y^m (p_m,h)$ is the $m$-meson yield spectrum at the altitude $h$. 
As the mesons are created in the hadronic interaction of cosmic rays 
and air nuclei,
the $Y^m (p_m,h)$ is written as
\begin{equation}
\label{eq:meson-yield}
Y^m (p_m,h) =
\int_{p_{proj}}\rho_{air}(h)~\cdot
\sum_i\sigma_{i}(p_{proj})\cdot\eta_i^m(p_{proj},p_m)\cdot\phi_i(p_{proj},h)
~dp_{proj}~,\\
\end{equation}
where $\rho_{air}(h)$ is the nuclear density of the air at altitude $h$,
$i$ stands for the kind of projectile ($p,\bar p,..,\pi^\pm$,..)  
for the hadronic interaction in the atmosphere,
$\sigma_{i}$ is the hadronic production cross section of the $i$ particle
and the air nuclei,
$\eta_i^m (p_{proj},p_m)$ is the $m$-meson production spectrum 
in the hadronic interaction of $i$ projectile and air nuclei,
and 
$\phi_{i}(p_{proj},h)$ is the momentum spectrum of $i$ particle at altitude $h$.
Note, we assume the superposition model for cosmic rays heavier 
than protons.

\begin{figure}[tbh]
\includegraphics[width=6.5cm]{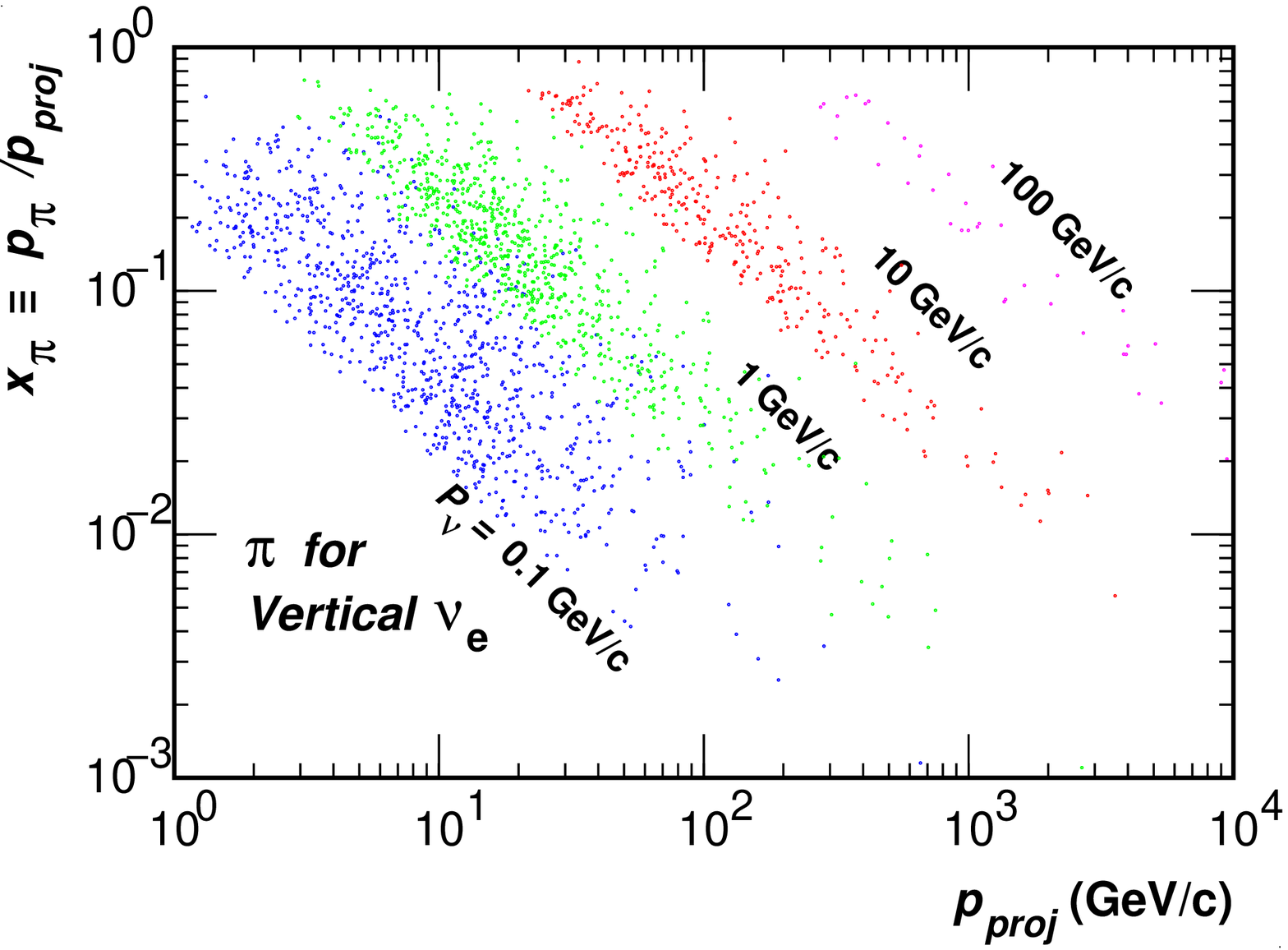}%
\includegraphics[width=6.5cm]{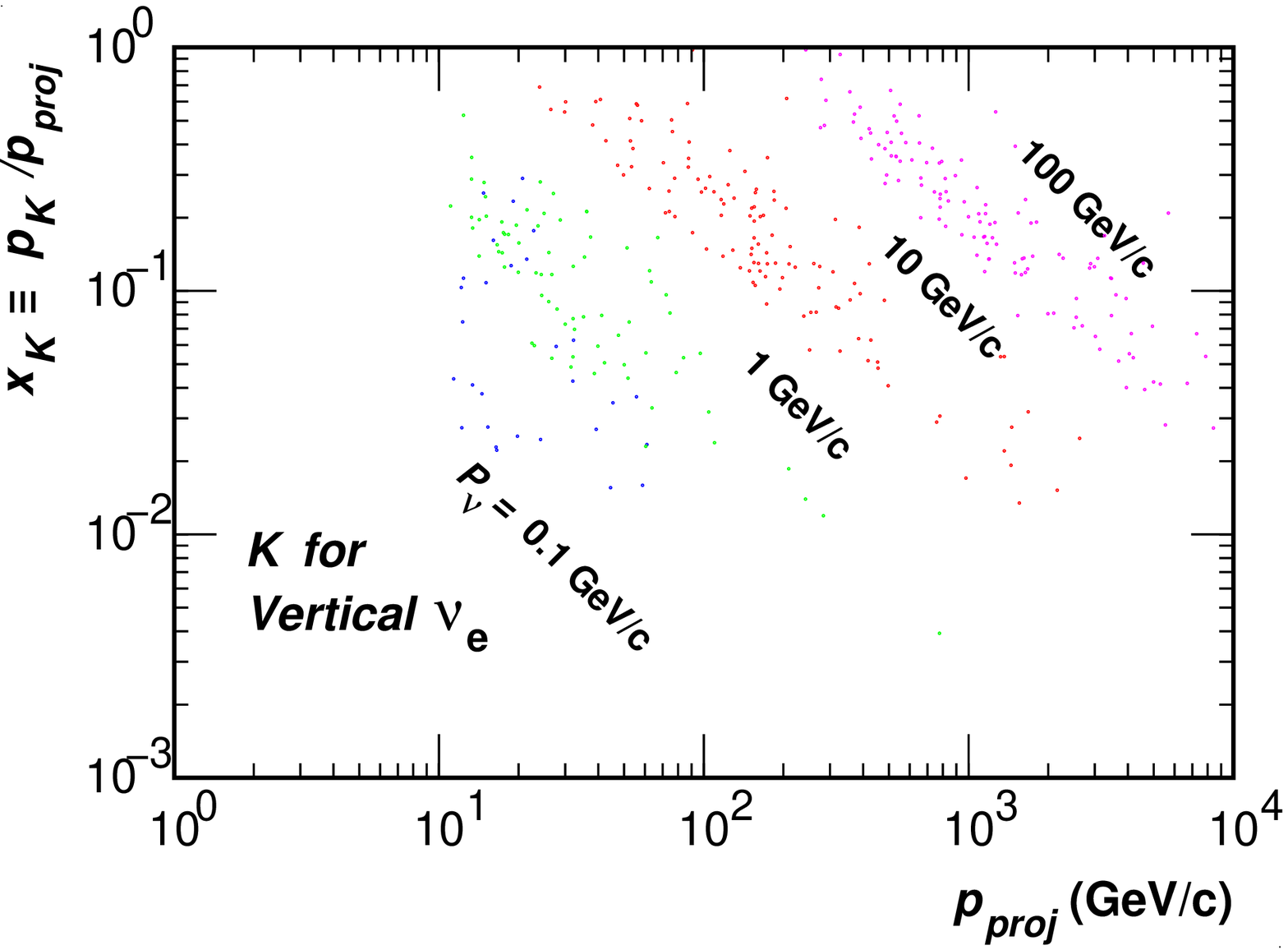}%
\\
\includegraphics[width=6.5cm]{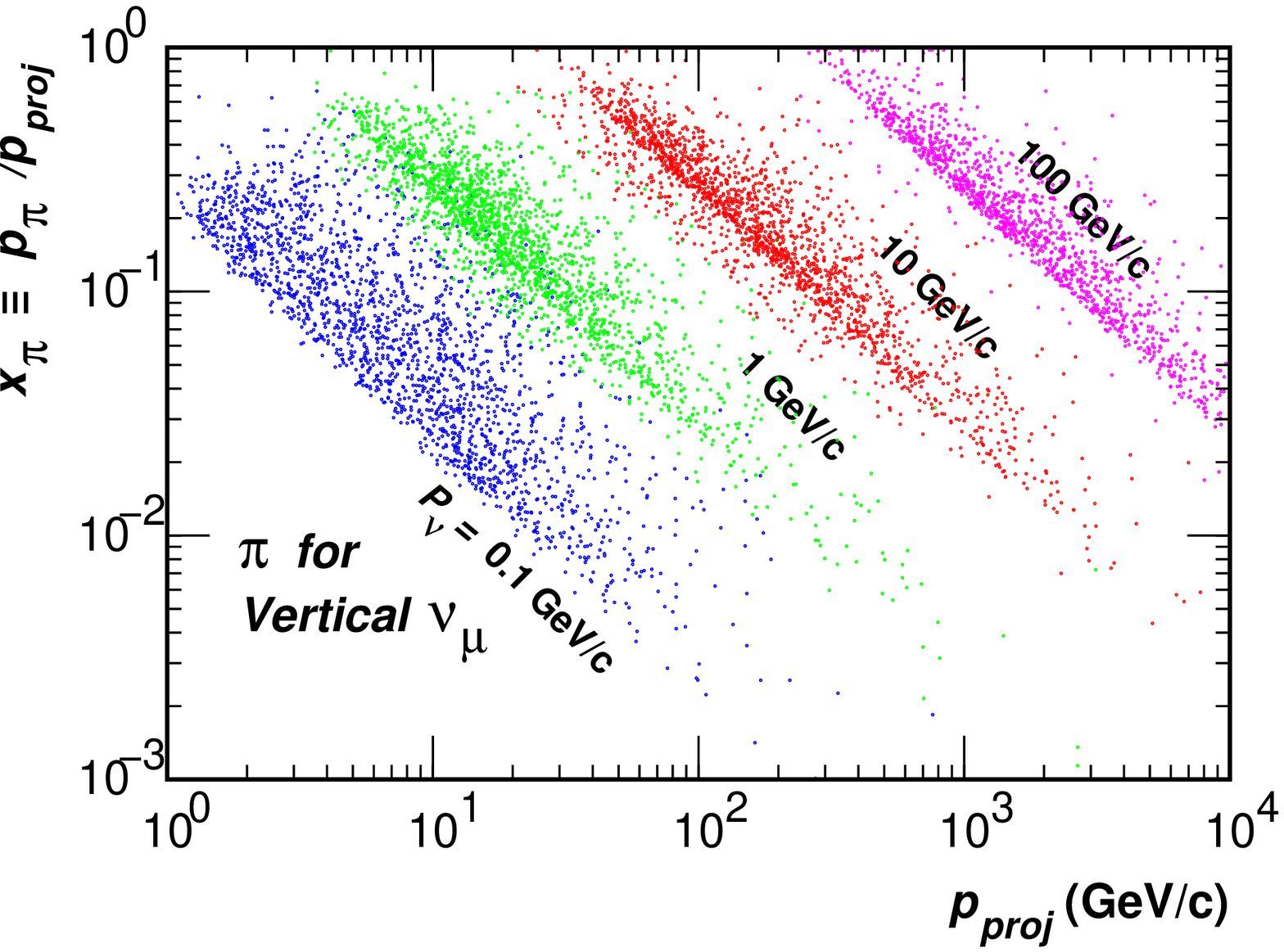}%
\includegraphics[width=6.5cm]{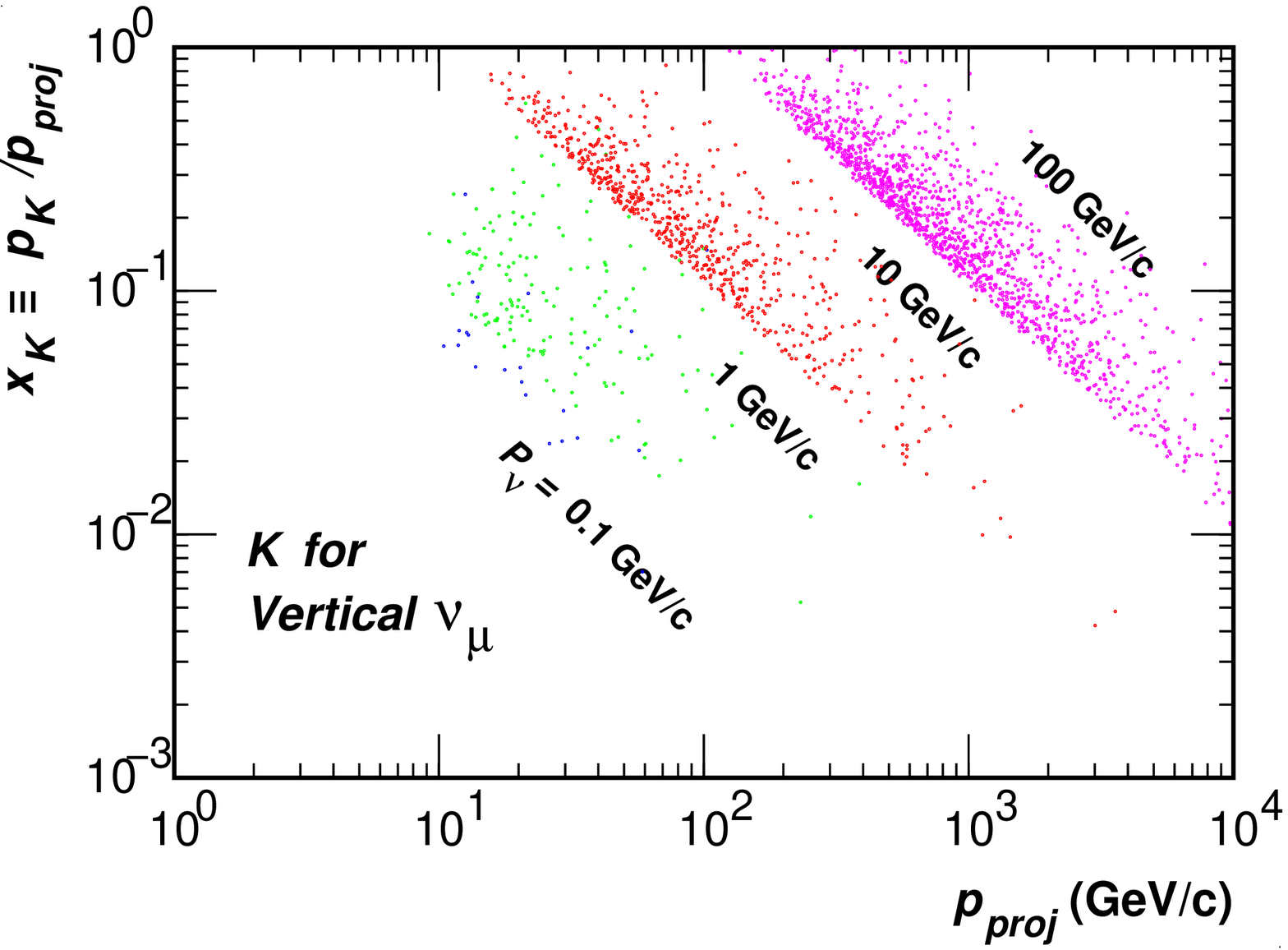}%
\\
\includegraphics[width=6.5cm]{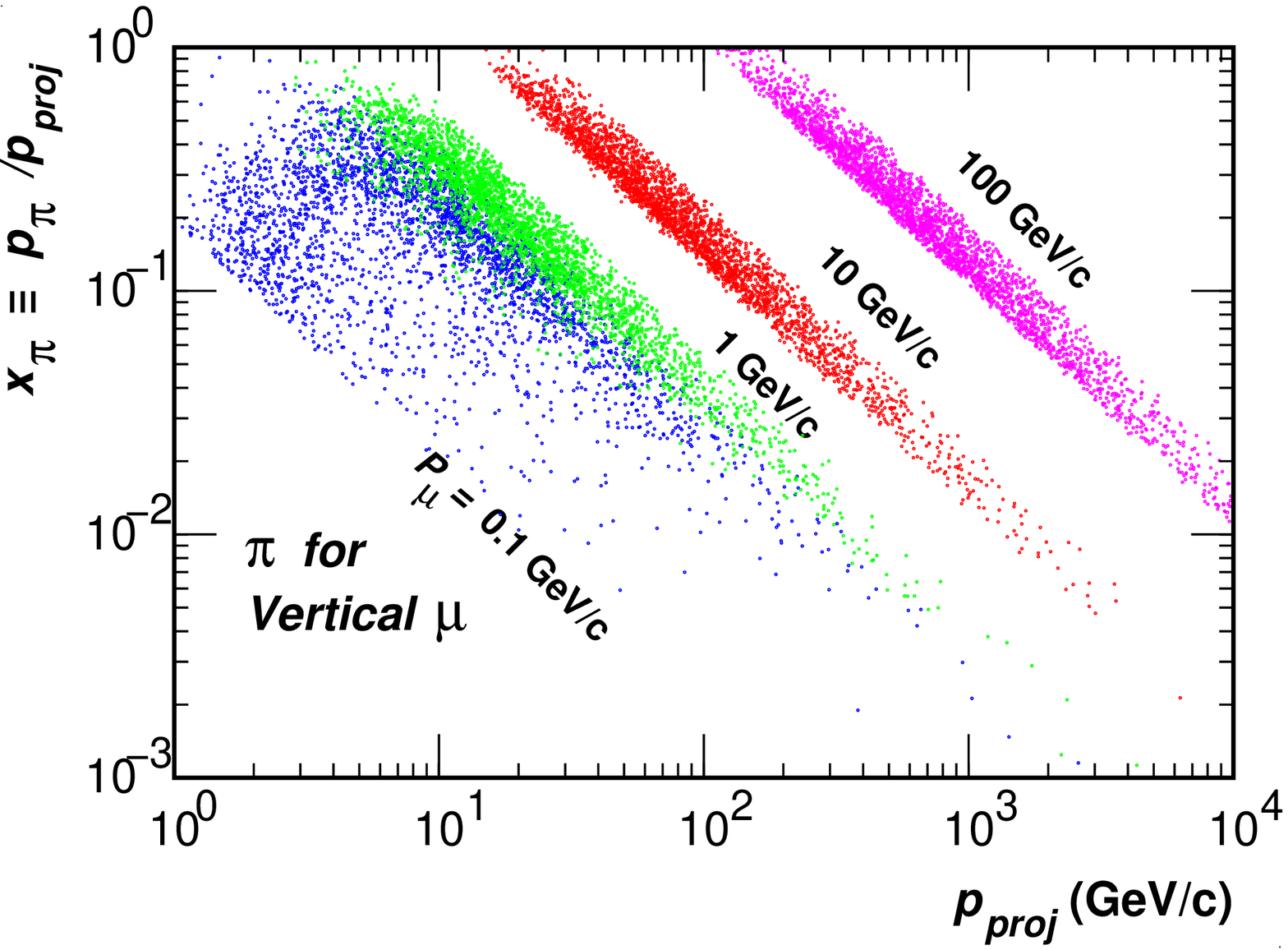}%
\includegraphics[width=6.5cm]{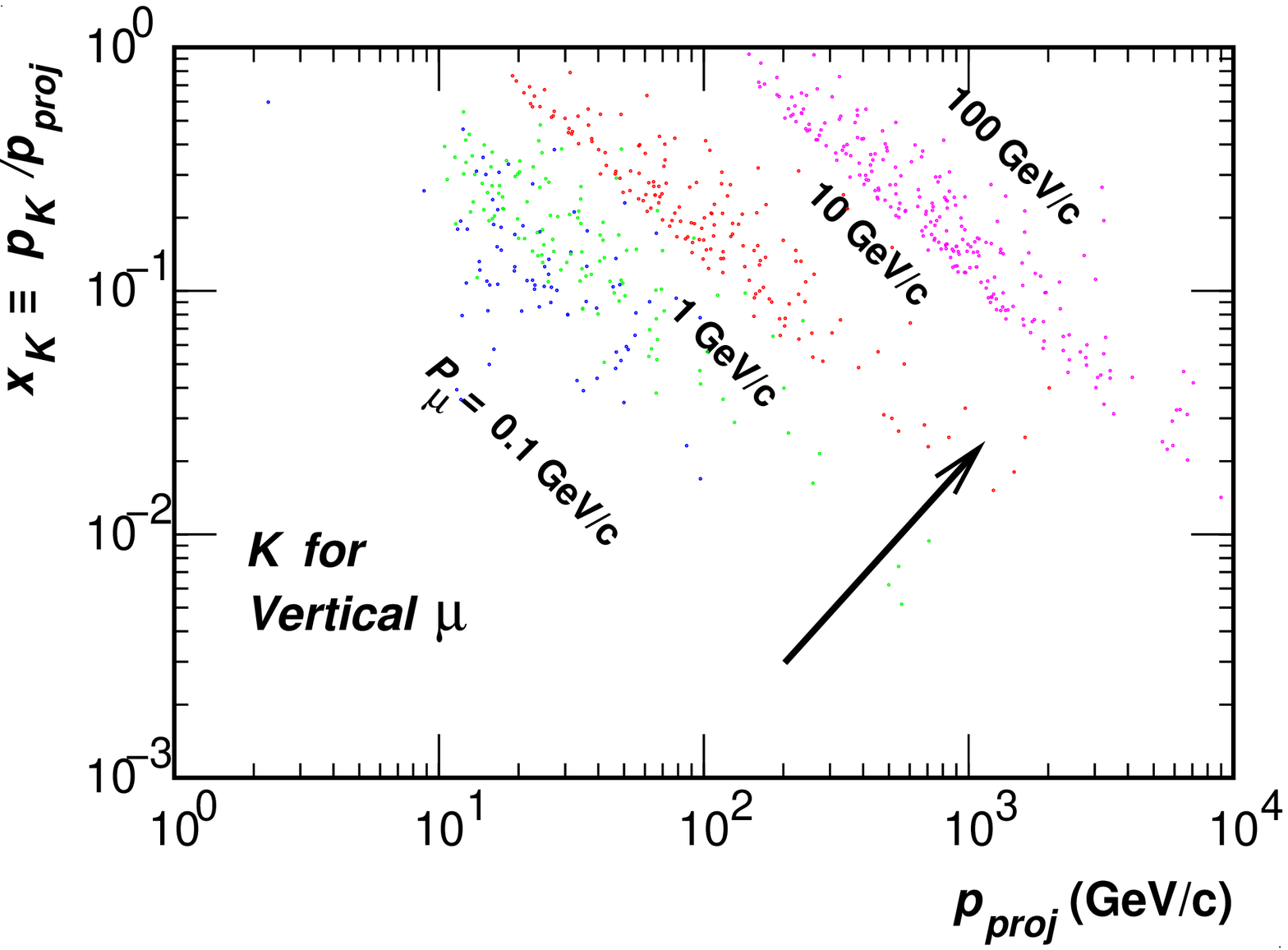}%
\caption{\label{fig:meson-scatter}
The scatter plot of $\pi$'s ($\pi^+$ and $\pi^-$, left) 
and $K$'s ($K^\pm$, $K^0_L$ and $K^0_s$, right) in 
the phase space ($p_{proj}-x$ plane) at their production
relevant to atmospheric $\mu$'s and $\nu$'s at the momenta,
0.1, 1.0, 10, and 100~GeV/c, at sea level for vertical directions.
Here, $x$ is defined as  $x \equiv p_m/p_{proj}$ in the rest frame of 
Air nuclei for $m=\pi$ or $K$.
We sampled 3,000 $\mu$'s and 3,000 $\nu$'s ($\nu_e$+$\nu_\mu$) at each momentum 
from the HKKM04 calculation.
The arrow in the right bottom panel shows the directions to which
$p_{\pi,K}$ increase.
%
}
\end{figure}

Substituting Eq.~\ref{eq:meson-yield} in Eq.~\ref{eq:meson-lepton}
and changing the integration order,  
we obtain
\begin{equation}
\label{eq:pm-int}
\phi_l(p_l) = 
\sum_{m}
\int_{p_m}\int_{p_{proj}}
\Big[
\int_h T_m^l(p_m, p_l, h) 
\rho_{air}(h) 
\sum_{i} \sigma_{i}(p_{proj})\eta_{i}^m (p_{proj},p_m)\phi_{i}(p_{proj},h) dh
\Big]
dp_{proj} dp_m~,
\end{equation}
The term inside the square brackets in Eq.~\ref{eq:pm-int} 
is interpreted as the contribution density in meson production 
phase spaces ($p_{proj}-p_m$  plane) to the lepton flux.
Using the scaling variable $x \equiv p_m/p_{proj}$ defined in the
rest frame of Air nuclei,
we show the contribution density calculated by the Monte Carlo 
calculation as a scatter plot in $p_{proj}-x$ plane 
(Fig.~\ref{fig:meson-scatter}).
Note, the variable $x$ is defined in the rest frame of Air nuclei.

Here, 
we sampled 3,000 $\mu$'s ($\mu^+ +\mu^-$) and 3,000 $\nu$'s 
($\nu_\mu+\bar\nu_\mu + \nu_e+\bar\nu_e$)
at 0.1, 1.0, 10, and 100~GeV/c 
from the HKKM04 calculation for atmospheric $\mu$ and $\nu$ fluxes
for vertical directions at sea level.
Then, we plotted the momenta of parent mesons,
$\pi$'s ($\pi^+$ and $\pi^-$) or $K$'s ($K^\pm$, $K^0_L$ and $K^0_s$),
and projectiles in Fig.~\ref{fig:meson-scatter} as a scatter plot.
Note, here and in the following discussions in this section, 
we do not distinguish the particles from antiparticles.

\begin{figure}[tbh]
\includegraphics[height=9cm]{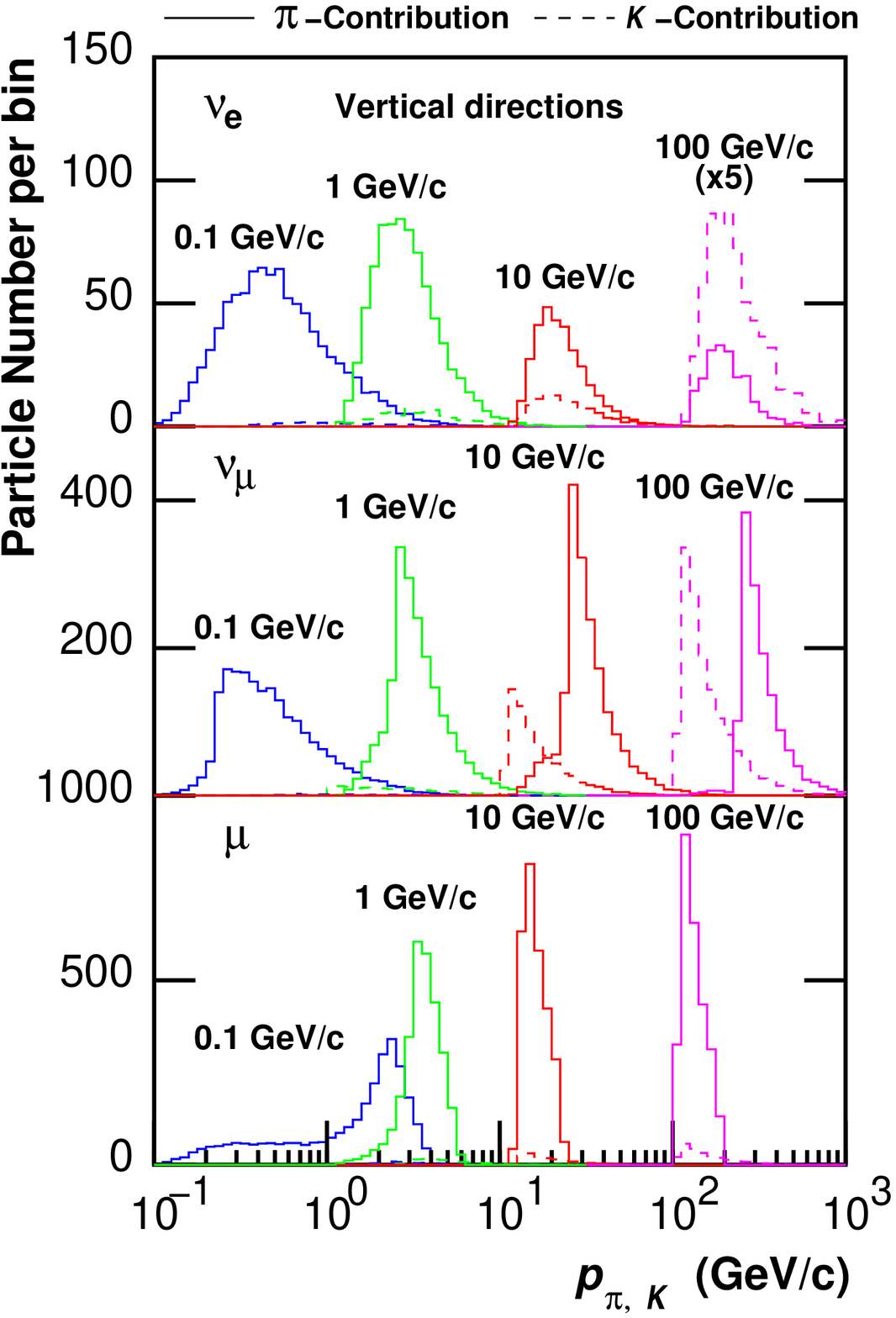}%
\hspace{5mm}
\includegraphics[height=9cm]{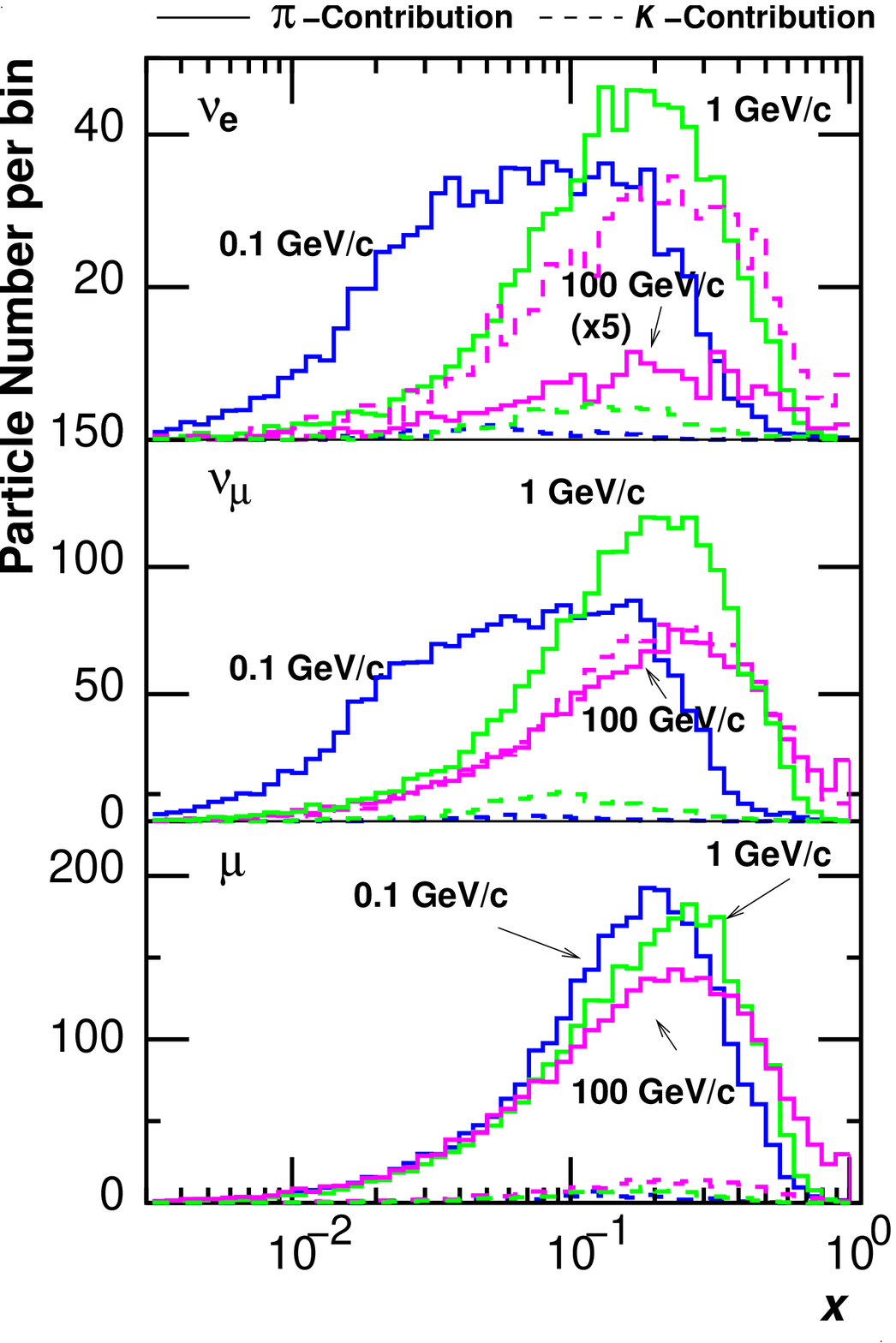}%
\caption{\label{fig:meson-hist}
Left panel: the momentum distributions of $\pi$'s and $K$'s relevant to 
the atmospheric $\mu$'s and $\nu$'s with fixed momenta,
0.1, 1.0, 10, 100~GeV/c, at sea level for vertical directions.
Right panel: the corresponding $x$ distributions for all except 
$p_{\mu,\nu}=$~10 GeV/c.
In these figures, total number of $\mu$'s or $\nu$'s ($\nu_e$+$\nu_\mu$) 
is normalized to 3,000, but the histograms are multiplied by a
factor of 5 for $\nu_e$ at 100~GeV/c.
}
\end{figure}

For the $\mu$'s, $\nu_\mu$'s and $\nu_e$'s, 
most $\pi$'s or $K$'s are 
concentrated in narrow stripes for each momentum above 1~GeV/c. 
In Fig.~\ref{fig:meson-hist}.
we project the points to the $\pi$ or $K$ momentum axis (left)
and to the $x$ axis (right) and show them in histograms. 
The total number of $\mu$'s or $\nu$'s ($\nu_e$+$\nu_\mu$)
is normalized to 3,000 for each momentum as in the scatter plot 
(Fig.~\ref{fig:meson-scatter}).
However, the histograms for the $\nu_e$ at 100 GeV/c are 
multiplied by a factor of 5, due to the rapid decrease of $\nu_e$ at this momentum.
The projections to the meson momentum axis are narrow distributions with sharp peaks
both for $\pi$'s and $K$'s,
which is proportional to
the integrand of the $p_m$ integration in Eq.~\ref{eq:pm-int}.
The projections to $x$ axis are very much like each other 
for all the $\mu$'s, $\nu_\mu$'s, and $\nu_e$'s above 1 GeV/c.
The projections to $x$ axis at 10 GeV/c are not shown in the figure,
since they are almost the same to those at 1 GeV/c or 100 GeV/c.

\begin{figure}[tbh]
\centerline{
\includegraphics[width=6.5cm]{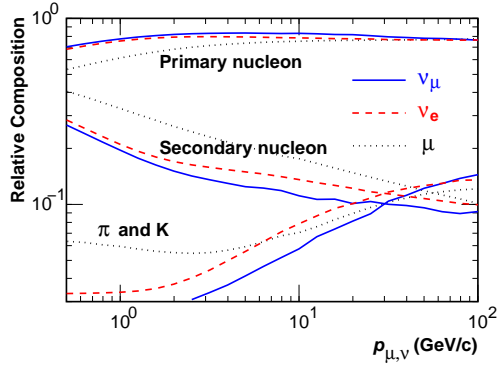}%
}
\caption{\label{fig:n2n}
The relative composition of the projectile of the last hadronic interaction,
for atmospheric $\mu$, $\nu_\mu$, and $\nu_e$.
}
\end{figure}

Changing the integration variables from $dp_{proj} dp_m $ to $(p_m/x^2)dxdp_m$,
and exchanging the integration order,
Eq.~\ref{eq:pm-int} can be rewritten as,
\begin{equation}
\label{eq:xf-int0}
\phi_l(p_l) = 
\sum_{m}
\int_{x}
\Big[
\int_{p_m}
\int_h 
T_m^l(p_m, p_l, h) 
\rho_{air}(h) 
\sum_i\{
\sigma_{i}(\frac{p_m}{x})
\cdot \eta_{i}^m(\frac{p_m}{x},p_m)
\cdot\phi_{i}(\frac{p_m}{x},h)\} dh
\frac {p_m} {x^2} dp_m 
\Big]
dx~.
\end{equation}
The projection to the $x$ axis is proportional to the integrand 
of the $x$ integration, and it is directly connected 
to the hadronic interaction model.
To illustrate this we introduce some assumptions.
First, we assume all the projectiles are nucleons.
In Fig.~\ref{fig:n2n}, 
we plot the relative composition of the projectiles 
for the interactions in which the parent meson of the leptons are 
created in the HKKM04 calculation.
We find the major projectiles are nucleons below 100 GeV/c, 
and the contribution of meson projectile remains $\lesssim$15\% even 
at higher momenta.

Next,
as the width of the $p_m$ distributions for fixed $p_l$ is narrow in the
left panel of Fig.~\ref{fig:meson-hist}, 
we approximate it with the $\delta$ function 
(zero width approximation) and 
write $T_m^l(p_m, p_l, h)$ as,
\begin{equation}
\label{eq:zero-width}
T_m(p_m, p_l, h)=\bar T_m^l(p_m, h)\cdot \delta(p_m - P_{ml}(p_l, h)),
\end{equation}
where $p_m = P_{ml}(p_l, h)$ is the average relation between
$p_m$, the momentum of mesons at altitude $h$, 
and $p_l$, the momentum of leptons at ground.
Then the $p_m$ integration is easily carried out and 
Eq.~\ref{eq:xf-int0} is rewritten as,
\begin{equation}
\label{eq:xf-int}
\phi_l(p_l) = 
\sum_{m}\int_{x}
\int_h \bar T_m^l(p_m, h) \rho_{air}(h) \sigma_{N}(\frac{p_m}{x})
\cdot\phi_{N}(\frac{p_m}{x},h)
\cdot 
\eta_{N}^m(\frac{p_m}{x},p_m) dh
\frac {p_m} {x^2}
dx~, %
\end{equation}
with $p_m = P_{ml}(p_l, h)$.
Note, 
as we are working with the flux sum of particles and anti-particles 
in this section, we introduced the iso-symmetric $m$ production function: 
\begin{equation}
\eta_{N}^m(p_{proj},p_m) \equiv (\eta_{p}^m (p_{proj},p_m) + \eta_{n}^m (p_{proj},p_m))/2~,
\end{equation}
for nucleons.
Note, the argument $h$ in the $P_{ml}$ is introduced to account for the
energy loss of $\mu$ for the leptons which are produced by $\mu$ decay.
The energy loss of the mesons before decay is very small.
The variation of the production height is estimated as 
$\sim\pm$~100 g/cm$^2$,
since the mean free path of the cosmic rays is 100~g/cm$^2$.
Therefore, the variation of $p_m$ due to the variation of production height
is $\sim \pm$0.2~GeV/c from the average, 
which is sufficiently small for leptons above 1~GeV/c.
We can write the relation of $p_m$ and $p_l$ in a much simpler 
function as $p_m = P_{ml}(p_l)$ without any altitude dependence for $p_l \gtrsim 1$~GeV/c.
Then $\eta_{N}^m({p_m}/{x},p_m)$ comes out of the $h$ integration in 
Eq.~\ref{eq:xf-int}, and it is rewritten as
\begin{equation}
\label{eq:xf-int2}
\phi_l(p_l) = \sum_{m}
\int_{x}
\Big[
\int_h \bar T_m^l(p_m, h) \rho_{air}(h) \sigma_{N}(\frac{p_m}{x})
\cdot\phi_{N}(\frac{p_m}{x},h)dh
\Big]
 \cdot
\Big[
\eta_{N}^m(\frac{p_m}{x},p_m) 
\frac{p_m}{x^2}
\Big]~ dx~,
\end{equation}
for $p_l \gtrsim 1$~GeV/c.
Now, the projection of contribution density to the $x$ axis is expressed 
by the product of two terms. One stands for the hadronic interactions,
and the other for the rest.
For later convenience, we write the expression as 
\begin{equation}
\label{eq:xf-int3}
\phi_l(p_l)
= \sum_{m}\phi_{l(m)}(p_l)
= \sum_{m}\int_{x}H_m^l(p_m, x)dx
~,
\end{equation}
with $p_m = P_{ml}(p_l)$ 
for $l=\mu$, $\nu_\mu$, and $\nu_e$ and $m=\pi^\pm,K^\pm,\cdots$ .
We will come back to the validity of the zero width approximation later.

\begin{figure}[tbh]
\includegraphics[width=6cm]{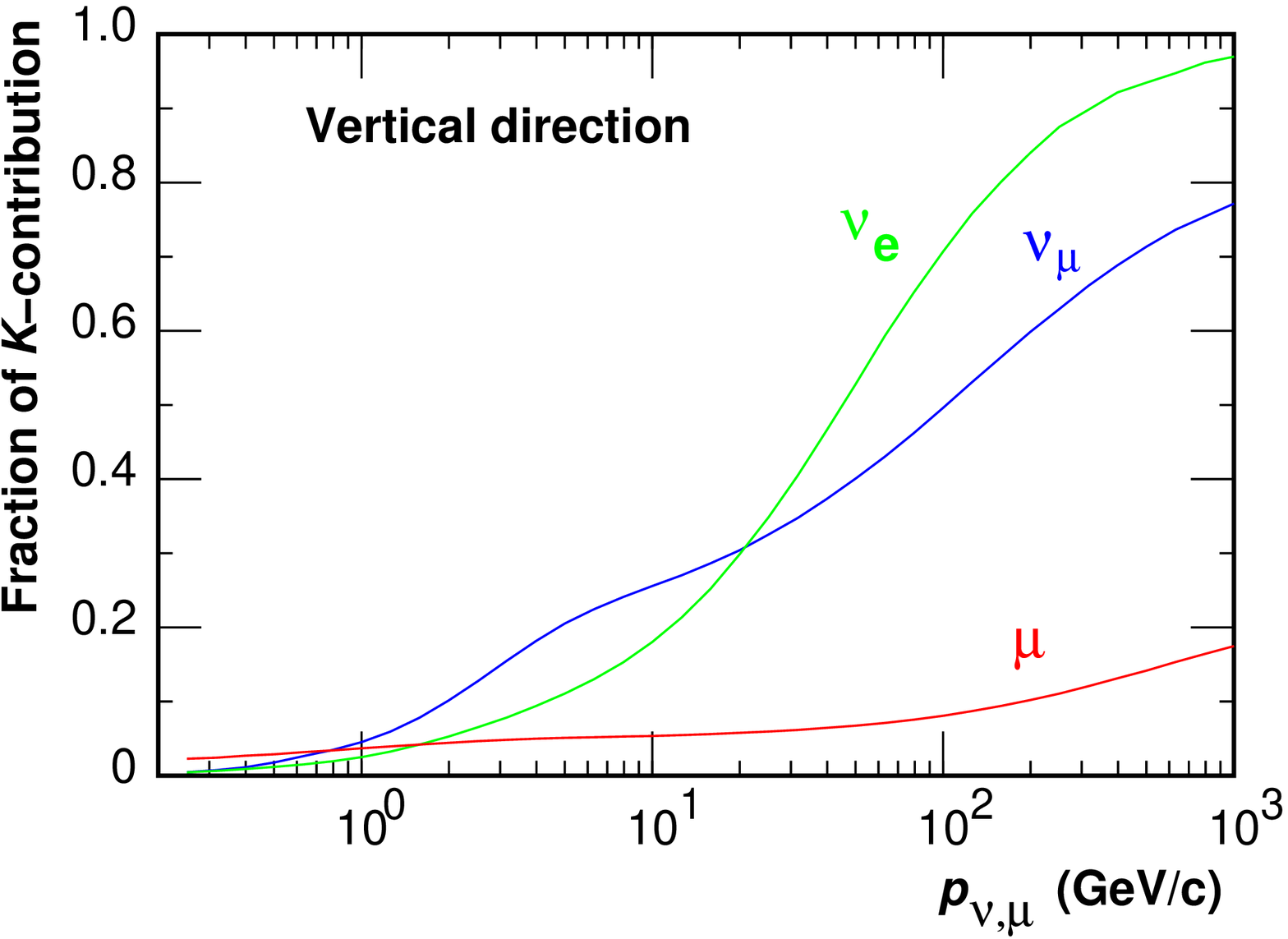}%
~~
\includegraphics[width=6cm]{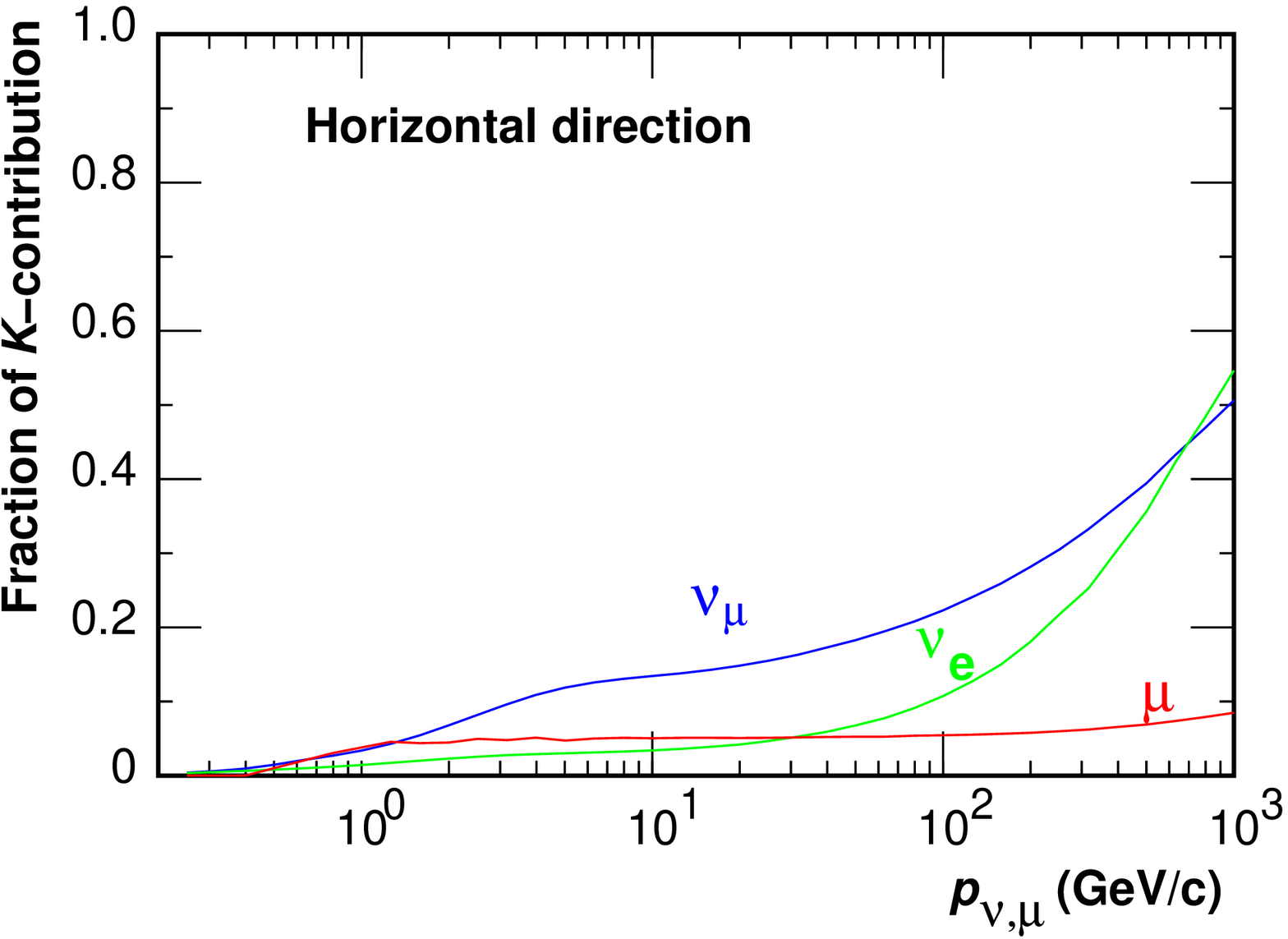}%
\caption{\label{fig:k-contrib}
The $K$-contribution to the vertical $\mu$'s and $\nu$'s (left), 
and to the horizontal $\mu$'s and $\nu$'s (right). 
}
\end{figure}

Here, we note it is difficult to study the hadronic interactions
for $K$ productions with atmospheric $\mu$ fluxes.
In Fig.~\ref{fig:k-contrib}, we depicted the contribution of $K$'s to the
atmospheric $\mu$'s and $\nu$'s in the ratio to the total flux as the functions 
of momentum.
The $K$-contribution is limited to the atmospheric $\mu$'s below 1~TeV/c,
while that the $K$-contribution to atmospheric $\nu$ is sizable above 10~GeV/c,
and is dominant above 100~GeV/c for vertical directions.
However, as most atmospheric $\mu$'s are produced in the $\pi$-decays,
we can use them to study the $\pi$ productions in the hadronic 
interactions, which are important in the calculation of atmospheric 
$\nu$ flux below 100~GeV/c.

\begin{figure}[tbh]
\centerline{
\includegraphics[width=6.5cm]{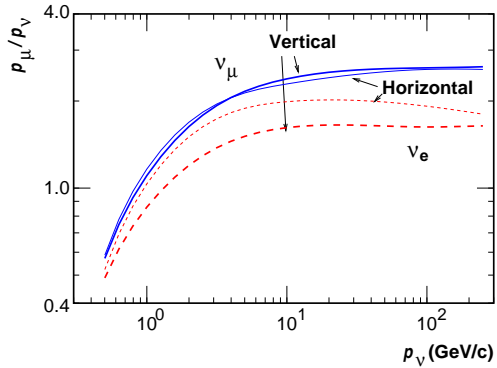}%
}
\caption{\label{fig:mu-nu-resp}
The momentum ratio of $\mu$'s to $\nu_\mu$'s or $\nu_e$'s,
whose parent $\pi$ momenta are the same on average.
The ``Horizontal'' in the figure means 
the momentum ratio between vertical $\mu$'s 
and horizontal $\nu_\mu$'s or $\nu_e$'s.
}
\end{figure}

We continue the study of hadronic interactions concentrating 
on the $\pi$ productions. 
In the following, we compare the variations of the different lepton fluxes
due to the error in the $\pi$ production of the hadronic interaction model.
This comparison should be done 
at the momentum where the parent $\pi$'s momenta are the same.
Therefore, the momentum relation $p_\pi=P_{\pi l}(p_l)$ is necessary 
for this study.
We construct the function from the Monte Carlo data, 
averaging the parent $\pi$'s momenta for a fixed lepton momentum.
However, 
the comparison is carried out among the fluxes of different kinds of leptons.
Therefore, we show the momentum relation between
$\mu$'s and $\nu_\mu$'s, and $\mu$'s and $\nu_e$'s
as ratios, namely
$P_{\pi \mu}^{-1}(P_{\pi \nu_\mu}(p_{\nu_\mu}))/p_{\nu_\mu}$ and  
$P_{\pi \mu}^{-1}(P_{\pi \nu_e}(p_{\nu_e}))/p_{\nu_e}$,
in Fig.~\ref{fig:mu-nu-resp}.
The ratio for horizontal directions is taken between horizontal
$\nu_\mu$ or $\nu_e$ and vertical $\mu$.
Note, the average momentum of parent $\pi$ for $\mu$'s are 
limited to $\gtrsim$~2~GeV/c,
since it is difficult for $\mu$'s produced in the decay of $\pi$'s with 
lower momenta than this limit to reach the ground level
(see Fig.~\ref{fig:meson-hist}). 
Therefore, there are no corresponding $p_\mu$ for $p_{\nu_\mu}$ or 
$p_{\nu_e} \lesssim$ 0.5 GeV/c. Note,
the relation depends on $\pi$'s energy spectrum at decay, 
therefore, 
on the primary cosmic ray spectra and interaction model.
However, changes in these do not affect the relations
greatly.

Let us consider that there are 2 interaction models with
the hadronic $\pi$ production function $\eta_{N}^\pi({p_\pi}/{x},p_\pi)$ 
and $\eta_{N}'^\pi({p_\pi}/{x},p_\pi)$,
and assume they are related by the factor function $\zeta({p_\pi}/{x}, x)$ as,
\begin{equation}
\eta_{N}'^\pi(\frac{p_\pi}{x},p_\pi) = \zeta(\frac{p_\pi}{x}, x)\cdot \eta_{N}^\pi(\frac{p_\pi}{x},p_\pi)~. 
\end{equation}
We will call $\eta_{N}'^\pi({p_\pi}/{x},p_\pi)$
the $\zeta$-modification of $\eta_{N}^\pi({p_\pi}/{x},p_\pi)$.
The $\zeta$-modified lepton fluxes are calculated as, 
\begin{equation}
\label{eq:modified-flux}
[\phi_{l(\pi)}(p_l)]_\zeta = 
\int_{x}
\zeta(\frac{p_\pi}{x}, x) \cdot H_\pi^l(p_\pi, x) dx~,
\end{equation}
where $p_\pi=P_{\pi l}(p_l)$ for $l=\mu$, $\nu_\mu$, and $\nu_e$.
Note, the error of the $\pi$ production function 
is also considered as a modification of the $\pi$ production function
of the perfect interaction model.

Using Eq.~\ref{eq:modified-flux} and 
the results of the Monte Carlo calculation partly shown in 
Fig.~\ref{fig:meson-hist},
it is now easy to study the effect of an error in the 
$\pi$ production function on the fluxes of $\mu$, $\nu_\mu$, and $\nu_e$,
with test functions for $\zeta$.
We assume that the modification function is expanded 
in the 2nd order B-spline functions~\cite{b-spline} as
\begin{equation}
\label{eq:vari-func}
\zeta(\frac{p_\pi}{x},x) = 1 + \sum_i C_{\zeta,i}(\frac{p_\pi}{x})\cdot\xi_i(\log(x)),
\end{equation}
where the 2nd order b-spline functions use here are defined as,
\begin{equation}
\label{eq:b-spline}
\begin{array}{c l c}
\xi_i(u) &= \frac{1}{2}\times \Bigg\{
\begin{array}{c c l}
(\frac{u-u_i}\Delta +1.5)^2 {\hspace{0.5cm}} & &(- 1.5 \le \frac{u-u_i}\Delta <- 0.5)\\
1.5- 2(\frac{u-u_i}\Delta)^2 {\hspace{0.5cm}}& &(- 0.5 \le \frac{u-u_i}\Delta <~~0.5)\\
(\frac{u-u_i}\Delta -1.5)^2 {\hspace{0.5cm}} & &(~~0.5 \le \frac{u-u_i}\Delta <~~1.5)\\
\end{array}\\
& {\rm and}\\
\xi_i(u) &= 0. \hspace{4.3cm} (\frac{u-u_i}\Delta < -1.5,~{\rm or}~1.5 \le \frac{u-u_i}\Delta)~,\\
\end{array}
\end{equation}
with $u_i = u_0 + i \cdot \Delta_u$ for $u=\log(x)$.
The B-spline function is often used to approximate a general 
continuous function.
It has a compact value region and is normalized as $\sum_i \xi_i(u)=1$.
The differentiability is not important here.
Note, when a function is approximated with the expansion of the B-spline functions,
the variations quicker than $\Delta$ are suppressed.

Then, Eq.~\ref{eq:modified-flux} can be rewritten as
\begin{equation}
\label{eq:modified-flux2}
[\phi_{l(\pi)}(p_l)]_\zeta = 
\int_{x}\big[1+\sum_i C_{\zeta,i}(\frac{p_\pi}{x}) \xi_i(\log(x))\big]
\cdot H_\pi^l(p_\pi, x)  dx~,
\end{equation}
and the difference to the original flux  as,
\begin{equation}
\label{eq:modified-flux3}
[\delta\phi_{l(\pi)}(p_l)]_\zeta \equiv[\phi_{l(\pi)}(p_l)]_\zeta -  \phi_{l(\pi)}(p_l)
= \sum_i C_{\zeta,i}(\frac{p_\pi}{x}) \cdot 
 \int_{x}\xi_i(\log(x))\cdot H_\pi^l(p_\pi, x) dx~.
\end{equation}
Therefore, the relative difference is calculated as,
\begin{equation}
\label{eq:modified-ratio}
\frac {[\delta\phi_{l(\pi)}(p_l)]_\zeta}{\phi_{l(\pi)}(p_l)}
= \frac{\sum_i C_{\zeta,i}(p_\pi/x) \int_{x}\xi_i(\log(x))\cdot H_\pi^l(p_\pi, x)dx}
{\int_{x} H_\pi^l(p_\pi, x)dx}
=\sum_i C_{\zeta,i}(\frac{p_\pi}{x}) \cdot R_i^l(p_\pi) 
\end{equation}
where, 
\begin{equation}
\label{eq:modified-ratio2}
R_i^l(p_\pi)
= {\int_{x}\xi_i(\log(x))\cdot H_\pi^l(p_\pi, x)dx}
\Big/
{\int_{x} H_\pi^l(p_\pi, x)dx}~,
\end{equation}
and $p_\pi=P_{\pi l}(p_l)$ 
for $l=\mu$, $\nu_\mu$, and $\nu_e$. 

The modification function corresponding to the error from the perfect $\pi$ 
production function is expected to be a slowly varying function of $x$
and not very different from 1 in all $x$ regions.
In the comparison of the $\pi$ production between 
an interaction model and the accelerator experiment data, 
we find typically differences of 20~$\sim$~30~\%
(see, for example, Fig.~15 of Ref.~\cite{Gaisser-Honda}).

\begin{figure}[tbh]
\centerline{
\includegraphics[width=6cm]{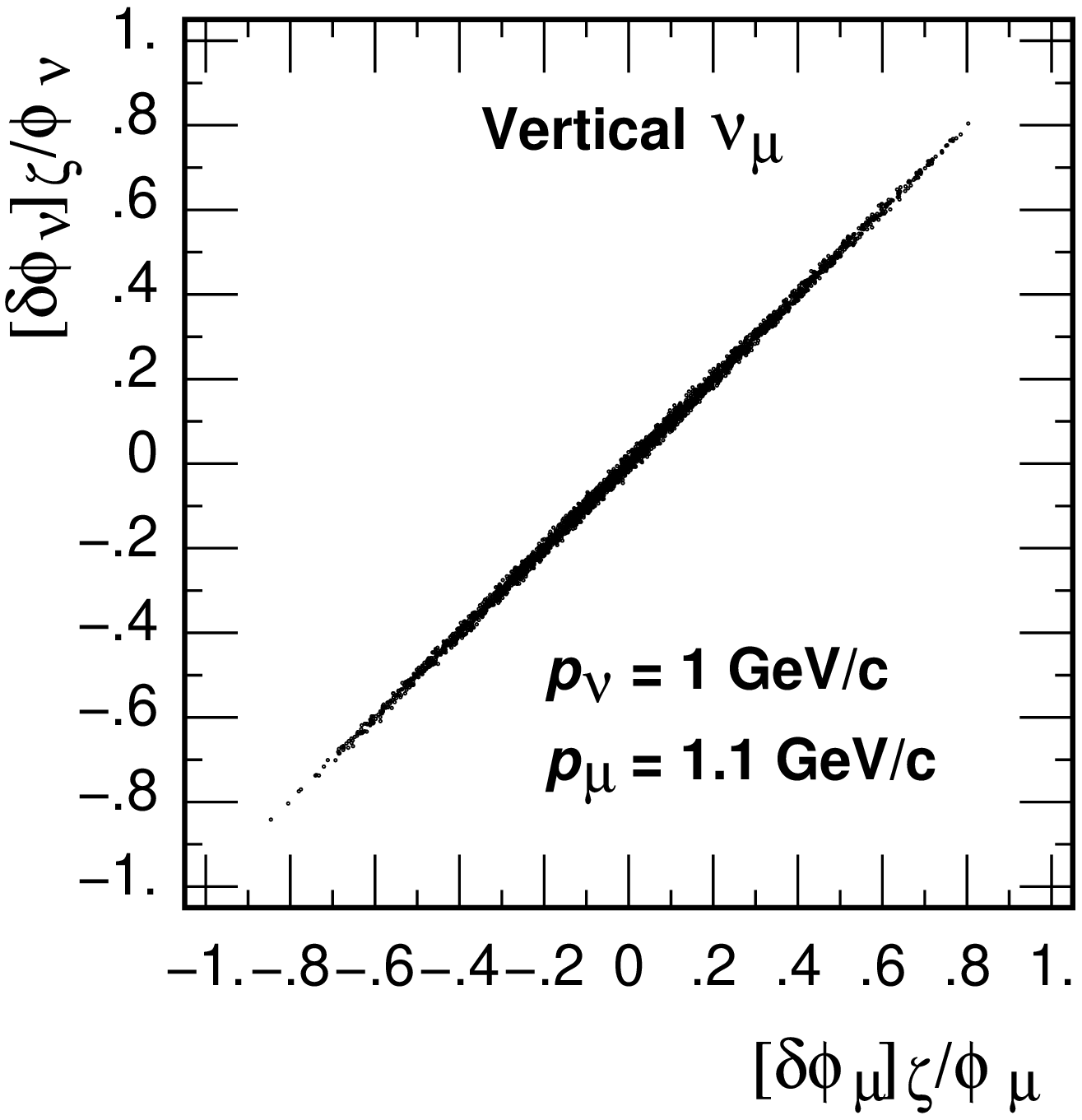}%
~~ 
\includegraphics[width=6cm]{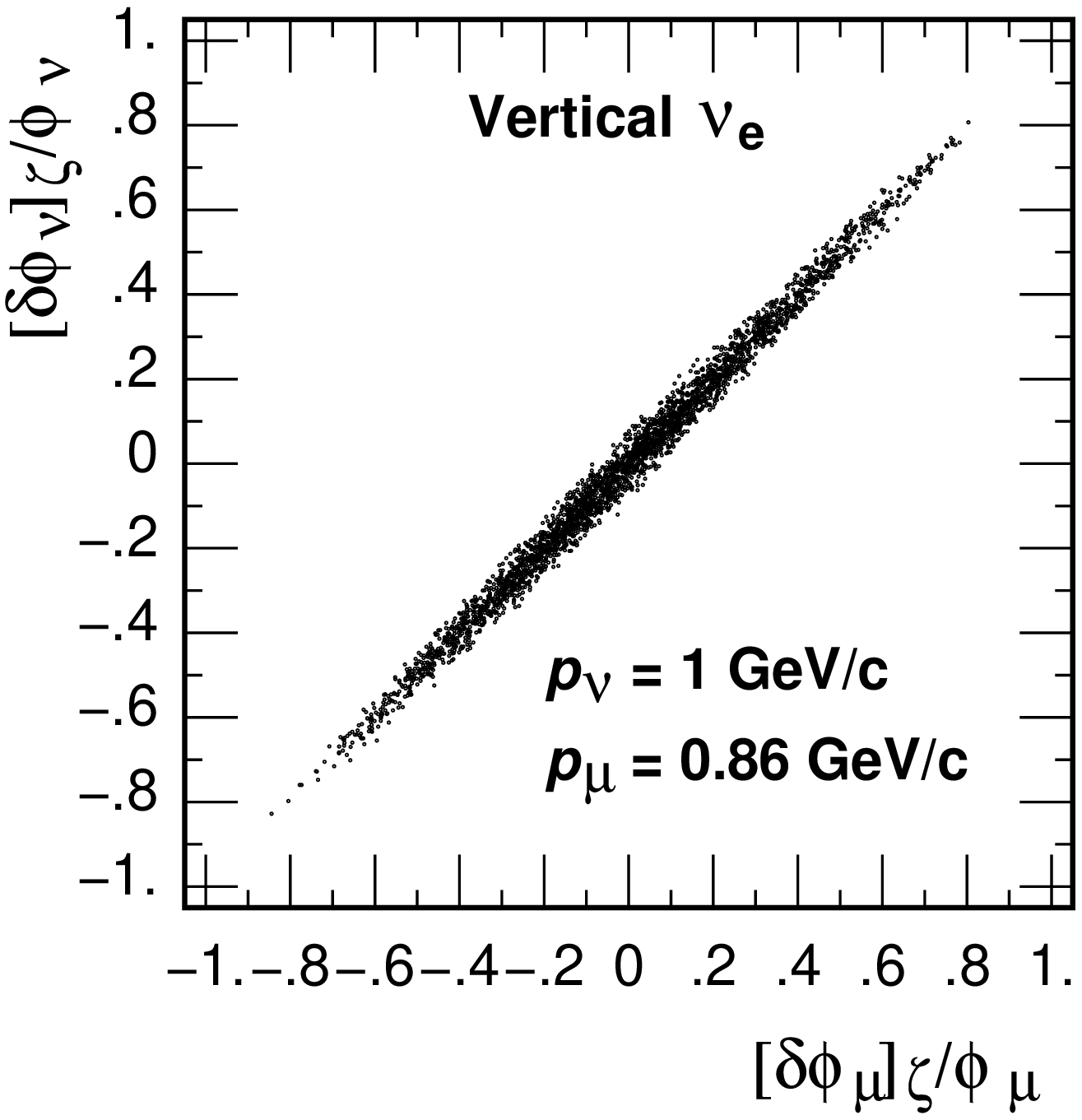}%
}
\caption{\label{fig:vari-scatter}
The scatter plots of 
($[\delta\phi_\mu]_\zeta/\phi_\mu$, $[\delta\phi_\nu]_\zeta/\phi_\nu$)
for $\nu_\mu$ (left) and $\nu_e$ (right).
The variation is calculated for 3,000 sets of random \{$C_{\xi i}$\} 
in [$-$1,1] 
for vertical directions at $p_\nu=$~1~GeV/c.
}
\end{figure}

Applying the Monte Carlo data to $H_\pi^l$ in Eq.~\ref{eq:modified-ratio2},
we can consider the artificial modifications of the $\pi$ productions function
with random numbers for 100~\% error in $\pi$ production function.
Taking a set of uniform random numbers in $[-1,1]$ for each $\{C_{\zeta,i}\}$,
we calculate the variation of
$[\delta\phi_\mu]_\zeta/\phi_\mu$, 
$[\delta\phi_{\nu_\mu}]_\zeta/\phi_{\nu_\mu}$ and
$[\delta\phi_{\nu_e}]_\zeta/\phi_{\nu_e}$, using the Eq.~\ref{eq:modified-ratio}
and $\Delta=0.5$ in Eq.~\ref{eq:b-spline}. 
The variations for 3,000 random $\{C_{\zeta,i}\}$ sets 
are plotted as a scatter plot in Fig.~\ref{fig:vari-scatter}.
Although the variations for the 
$[\delta\phi_\mu]_\zeta/\phi_\mu$, 
$[\delta\phi_{\nu_\mu}]_\zeta/\phi_{\nu_\mu}$ or
$[\delta\phi_{\nu_e}]_\zeta/\phi_{\nu_e}$ are large,
we find a narrow concentrations to the
$[\delta\phi_{\mu}]_\zeta/\phi_\mu = [\delta\phi_{\nu}]_\zeta/\phi_\nu$ line
there.

\begin{figure}[tbh]
\centerline{
\includegraphics[width=5cm]{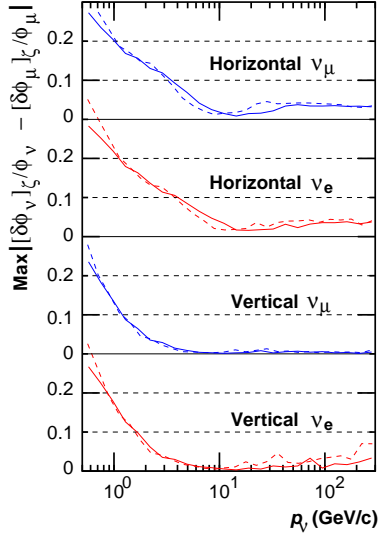}%
}
\caption{\label{fig:varimax}
The maximum of
$|[\delta\phi_{\nu_\mu}]_\zeta/\phi_{\nu_\mu}-[\delta\phi_\mu]_\zeta/\phi_\mu|$ 
and
$|[\delta\phi_{\nu_e}]_\zeta/\phi_{\nu_e}-[\delta\phi_\mu]_\zeta/\phi_\mu |$.
The solid line assumes  $\Delta=0.5$ for the B-spline functions (Eq.~\ref{eq:b-spline}),
and the dashed line $\Delta=0.25$.
They are calculated analytically using the Monte Carlo data.
}
\end{figure}

As $C_{\zeta,i}$ varies freely in $[-1,1]$, the maximum difference is calculated as,
\begin{equation}
\label{eq:varimax}
{\rm Max} \Big|\frac{[\delta \phi_{\nu(\pi)}(p_\nu)]_\zeta}{\phi_{\nu(\pi)}(p_\nu)}
- \frac{[\delta \phi_{\mu(\pi)}(p_\mu)]_\zeta}{\phi_{\mu(\pi)}(p_\mu)}\Big|
= {\rm Max} \Big|\sum_i C_{\zeta,i}\Big[R_i^\nu(p_\pi) -R_i^\mu(p_\pi)\Big]\Big|
=\sum_i \Big|R_i^\nu(p_\pi) -R_i^\mu(p_\pi)\Big|,
\end{equation}
where $\nu$ stands for $\nu_\mu$ and $\nu_e$.
In Fig.~\ref{fig:varimax}, we show the maximum difference between
$[\delta\phi_{\mu}]_\zeta/\phi_\mu$ and $[\delta\phi_{\nu_\mu}]_\zeta/\phi_{\nu_\mu}$, 
and
$[\delta\phi_{\mu}]_\zeta/\phi_\mu$ and $[\delta\phi_{\nu_e}]_\zeta/\phi_{\nu_e}$
as a function of $p_\nu$, and they are small for $p_\nu \gtrsim$~1~GeV/c.
Note, for the horizontal directions, the maximum difference 
between horizontal $\nu$'s and vertical $\mu$ is calculated.
If we assume 20~\% as the maximum error for the $\pi$ production function, 
$C_{\zeta,i}$ is sampled in $[-0.2,0.2]$ instead of $[-1,1]$.
Then, the maximum differences are multiplied by 0.2 to the values shown in 
Fig.~\ref{fig:varimax}.

For the modifications, or for the error of the hadronic 
$\pi$ production function,
we have the approximate relation 
\begin{equation}
\frac{[\delta \phi_\mu(p_\mu)]_\zeta}{\phi_\mu(p_\mu)}
\simeq 
\frac{[\delta\phi_{\nu_\mu}(p_{\nu_\mu})]_\zeta}{\phi_{\nu_\mu}(p_{\nu_\mu})}
\simeq \frac{[\delta\phi_{\nu_e}(p_{\nu_e})]_\zeta}{\phi_{\nu_e}(p_{\nu_e})}~,
\end{equation}
for the lepton flux whose parent momenta are the same, or
$P_{\pi\mu}(p_\mu)
=P_{\pi{\nu_\mu}}(p_{\nu_\mu})
=P_{\pi{\nu_e}}(p_{\nu_e})$.
Note, the approximate relation becomes invalid below $\sim$~1~GeV/c,
as is seen from Fig.~\ref{fig:varimax}.

The maximum difference with $\Delta=0.25$ for the
B-spline functions (Eq.~\ref{eq:b-spline}) is shown with a dashed 
line in Fig.~\ref{fig:varimax}. 
However, the maximum difference with $\Delta=0.25$ is almost the same as 
that with $\Delta=0.5$, implying $\Delta=0.5$ is fine enough in this study.

We have some comments for the zero width approximation.
This approximation is valid for the $\mu$'s
until the variation of the $\mu$-energy lost in the air becomes crucial.
This is because the energy of $\mu$ in $\pi$-decay is limited to
\begin{equation}
(1 - \beta \cdot \frac{m_\pi^2 - m_\mu^2}{m_\pi^2 + m_\mu^2})\cdot\bar E_{\mu}
\le E_{\mu}
\le
(1 - \beta \cdot \frac{m_\pi^2 - m_\mu^2}{m_\pi^2 + m_\mu^2})\cdot\bar E_{\mu}~,
\end{equation}
or approximately in $0.73\cdot \bar E_\mu \le E_\mu \le 1.27 \cdot\bar E_\mu$,
where
$\bar E_\mu = E_\pi\cdot({m_\pi^2 + m_\mu^2})/{2m_\pi}$  is the average 
energy of $\mu$ in $\pi$-decay.
However, the $E_{\nu_\mu}$ in the $\pi$-decay distributes uniformly in
$[0, ~2 \bar E_{\nu_\mu}]$, 
where $\bar E_{\nu_\mu}= E_\pi\cdot({m_\pi^2 - m_\mu^2})/{2m_\pi}$ 
is the average energy of $E_{\nu_\mu}$ in the $\pi$ decay.
For the decay products of the $\mu$'s, the 3-body decay phase space is convoluted. 
Therefore, a wide momentum distribution is expected for the momenta of parent 
$\pi$'s of $\nu$'s. 
Note, the $\mu$ spin effect is a minor effect in this discussion.
However, the steep $\pi$-decay spectrum makes the momentum distributions 
effectively narrower.
Most of the momenta of parent $\pi$'s distribute within 
$\bar p_\pi/2 \lesssim p_\pi \lesssim \bar p_\pi$,
where $\bar p_\pi$ is the average momentum of parent $\pi$'s.
Modifying the delta-function in Eq.~\ref{eq:zero-width} to 
a narrow distribution function of $p_m$ and retaining $p_m$ 
integration in the Eq.~\ref{eq:xf-int3},
we could carry out a more general study than that presented here.
However, we expect an almost the same result due the weak dependence 
of the $\eta_N^m(p_{proj},p_m)$ on $p_{proj}$.

In this section,
we have studied the response of atmospheric $\mu$ and $\nu$ fluxes 
to error in the $\pi$ production in the hadronic interaction model.
We have shown that a modification affects the atmospheric $\mu$
and $\nu$ fluxes originating from the $\pi$ decay at the same rate, 
namely
$\Delta\phi_\mu/\phi_\mu \simeq \Delta\phi_{\nu_\mu}/\phi_{\nu_\mu} 
\simeq \Delta\phi_{\nu_e}/\phi_{\nu_e}$, for $p_\nu \gtrsim$~1~GeV/c.
This is an important relation, 
since the error of the hadronic interaction model could be sensed by 
a comparison of calculated and observed $\mu$ flux data,
especially when accurately measured $\mu$ flux data are 
available.
The relation could be used not only to estimate 
the error in the calculation 
but also to tune the hadronic interaction model.
However, this is true only when we carry out the height integration in 
Eq.~\ref{eq:xf-int2} for $\mu$ and $\nu$ correctly.
In other words, the propagation of particles in the air must be 
carried out correctly (disregarding the hadronic interactions).
For an error of the physical input which works in the same 
direction for the atmospheric $\mu$ and $\nu$ fluxes, like an 
error in the primary cosmic ray flux model,
the uncertainty may be merged in the uncertainty of the
hadronic interaction model, and is calibrated by the atmospheric 
muon flux data collectively.
However, there are some physical inputs whose error works in different
directions for the atmospheric $\mu$ and $\nu$ fluxes.
We must be careful about such uncertainties.

\section{\label{sec:atmosphere}Atmospheric Density Profile and Interaction Cross-Section}

In this section,
we study the effects of error in the atmospheric density profile
and hadronic interaction cross section on the atmospheric muon and neutrino fluxes.
Both effects are relatively smaller than those of 
hadronic interaction and primary cosmic ray flux, 
but the errors work differently on the atmospheric muon flux and neutrino flux.
Therefore, it is important to estimate the error for the study of 
hadronic interaction, then for the calculation  
of the atmospheric neutrino flux.
Note, we treat the interaction cross section separately 
from the dynamics of hadronic interaction.

\begin{figure}[tbh]
\includegraphics[width=7cm]{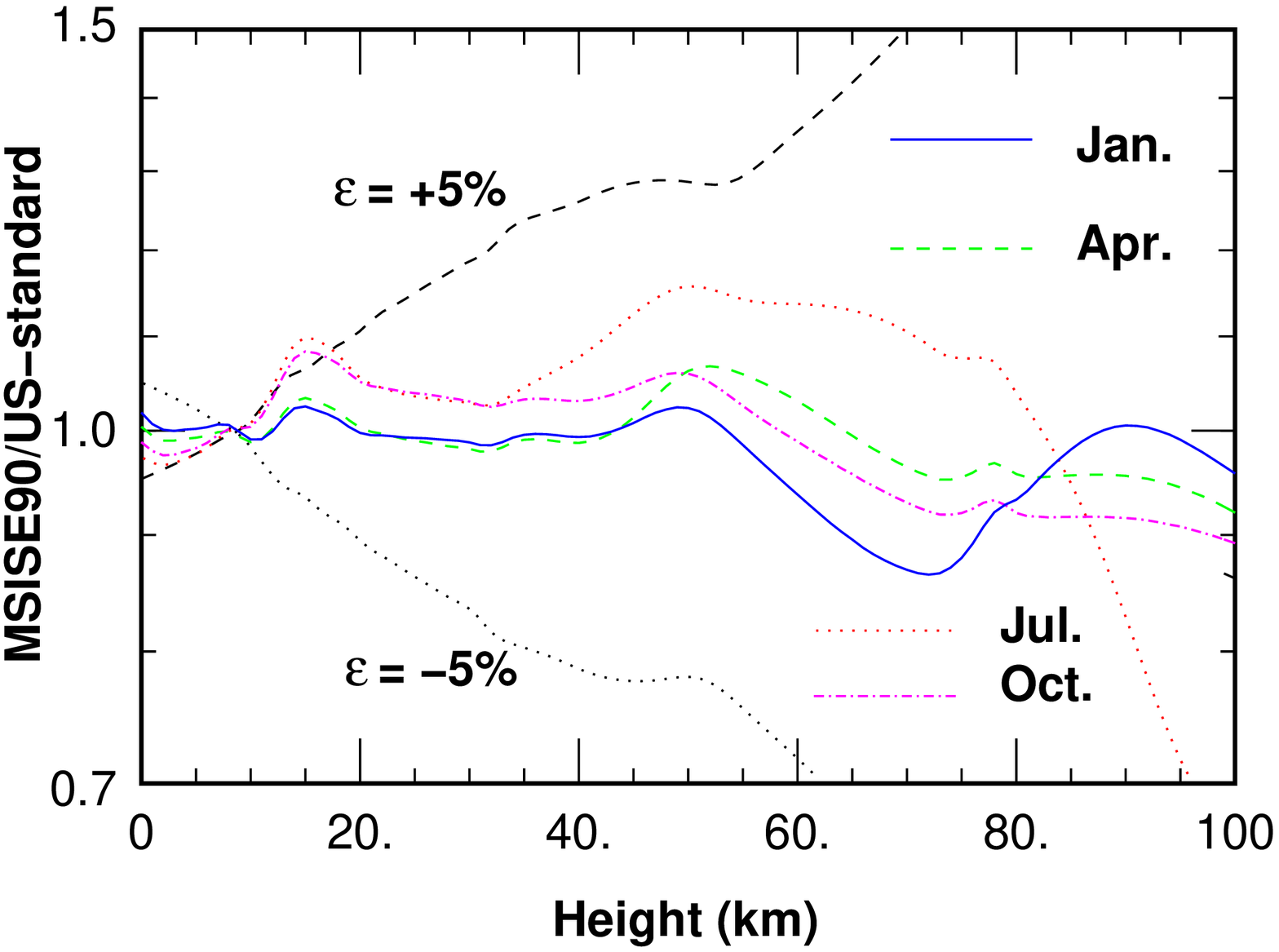}%
~
\includegraphics[width=7cm]{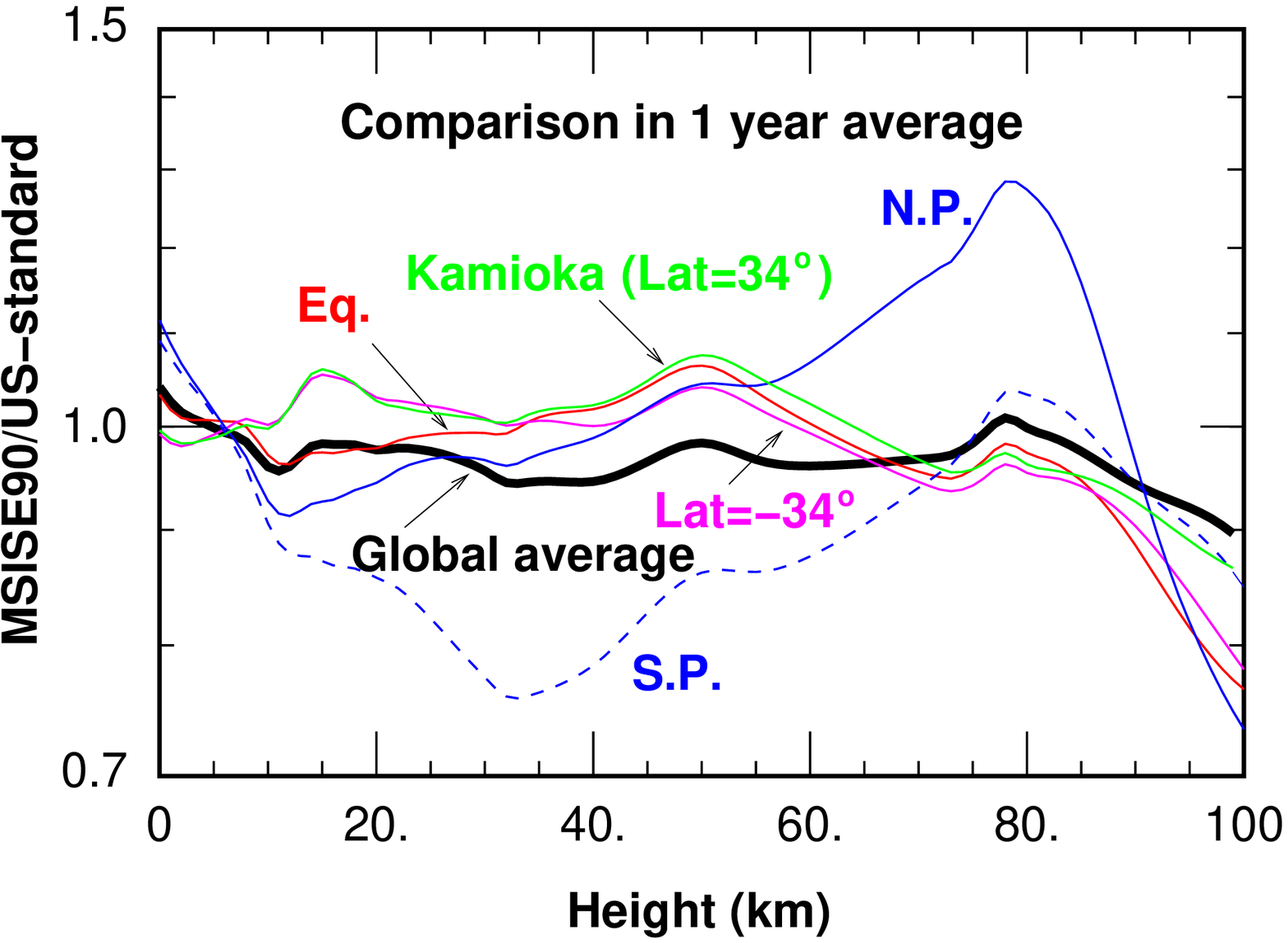}%
\caption{\label{fig:atmosphere}
Left panel: the comparison of air density profile of MSISE90\cite{msise90} 
and US-standard '76\cite{us-standard} in ratio at Kamioka in different seasons.
The variations corresponding to $\varepsilon=\pm5~\%$ in 
Eq.~\ref{eq:scaleheight} are also shown.
Right panel: the same as the left panel, but at different latitude in all season
average. The global average in the MSISE90 model is shown by the thick line.
}
\end{figure}

As the atmospheric density profile, 
the US-standard '76 atmosphere model~\cite{us-standard}
is generally used in the calculation of atmospheric neutrino flux,
including the HKKM04 calculation.
The density profile of the US-standard '76 atmosphere model is 
compared with that of newer atmosphere model MSISE90~\cite{msise90}
as a ratio in Fig.~\ref{fig:atmosphere}.
The MSISE90 is considered the more realistic atmosphere model, 
since it gives the position and time dependent atmospheric variations.
In the left panel,  
we show comparison of the atmospheric density profile at Kamioka for 
different seasons.
We find the maximum difference below 40~km is $\sim$~10~\% 
in summer (Jul.) and autumn (Oct.), but that the MSISE90 
air density profile is very close 
to that of US-standard '76 in winter (Jan.) and spring (Apr.).
In the right panel, we show the comparison of one-year-average at
different latitudes (Lat$=$ $-$90, $-$34, 0, 34, and 90),
and the global average with the US-standard '76.
The global average agrees well with the US-standard '76
within $\sim$ 5~\% except for very high latitude.

\begin{figure}[tbh]
\includegraphics[width=7cm]{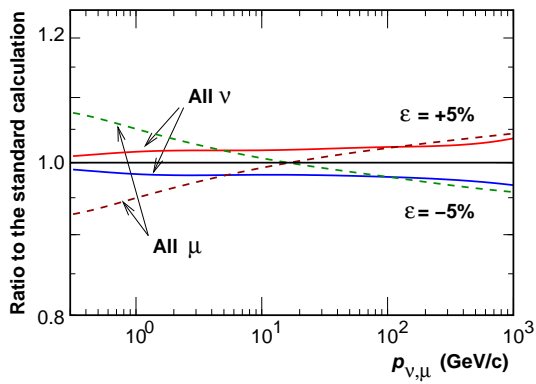}%
\caption{\label{fig:vari-by-air}
The variation of atmospheric muon and neutrino fluxes 
for a change of atmospheric density profile corresponding to the 
$\varepsilon\pm 5~\%$ in Eq.~\ref{eq:scaleheight} 
from the US-standard '76 atmosphere model. 
}
\end{figure}

To study the effect of the uncertainty of the air density profile,
we consider the modification of the US-standard '76 air density profile as
\begin{equation}
\label{eq:scaleheight}
\rho_{us,\varepsilon}(h) = \frac 1 {1+\varepsilon} \cdot \rho_{us}(\frac h {1+\varepsilon})~,
\end{equation}
where $\rho_{us}(h)$ is the atmospheric density profile of the US-standard '76,
and $h$ is the altitude.
Note the static solution for compressible gas in the gravitational field 
is expressed as 
\begin{equation}
\label{eq:static-air}
\rho(h) = \rho_0 \cdot e^{-\frac{h}{h_s}},
\end{equation}
and the scale height $h_s$ is proportional to the absolute temperature.
Therefore, the $\varepsilon$ in Eq.~\ref{eq:scaleheight}
corresponds to the change of the atmospheric temperature.
We consider the variation of $\varepsilon =\pm 5$~\% in 
Eq.~\ref{eq:scaleheight} for the seasonal variation.
Actually, the variation from winter--spring to summer--autumn is approximately 
the same as the variation of $\varepsilon=0 \sim +5~\%$  below 20~km 
(Fig.~\ref{fig:atmosphere}).
In Fig.~\ref{fig:vari-by-air},
we plot the variation of atmospheric muon and neutrino fluxes
as a ratio
for the variation of $\varepsilon= \pm 5~\%$ in Eq.~\ref{eq:scaleheight}.

Note, 
the variation of air density at the production height of $\pi$'s and $K$'s 
is the main reason for the variation of the fluxes of atmospheric neutrinos
and high energy muons.
The decay and the interaction are competitive processes for mesons 
($\pi$'s and $K$'s), and they balance when
[Flight Length Before Decay] = [Interaction Mean Free Path], or 
\begin{equation}
\label{eq:int-decay-valence}
c \tau \cdot E / Mc^2 =  A_{air} / \sigma_m \rho N_a
\end{equation}
is satisfied.
Here, $\tau$ is the lifetime of the particle,
$E$ is the energy and $M$ is the mass of the particle,
$A_{air}$ the average mass number of air nuclei,
$\sigma_m$ the interaction cross section of the meson and air nuclei,
$\rho$ the mass density of the air,
and $N_a$ the Avogadro constant.
The production height is approximated by the first interaction height 
of cosmic rays, i.e. $A_{air}/\sigma_{cr} N_a\sim$~100~g/cm$^2$
in the air depth,
where $\sigma_{cr}$ is the interaction cross section of cosmic rays.
In the US-standard '76 atmosphere model,
it is calculated as $\sim$ 16~km a.s.l. and the air density is 
$\sim 0.16\times 10^{-3}$~g/cm$^3$, for vertical directions.
The energies with which the decay and interaction balance are approximately
90~GeV for $\pi^{\pm}$, 170 GeV for $K_L^0$, and 690 GeV for $K^{\pm}$ there.
Below these energies, most $\pi$'s or $K$'s decay producing the
muons and neutrinos, and above these energies, most of them 
interact with the air nuclei, producing lower energy $\pi$'s or $K$'s.
From Eq. \ref{eq:scaleheight}, we find the air density at constant air depth
decreases for $\varepsilon > 0$ and increases for $\varepsilon < 0$.

The actual production height of $\pi$'s and $K$'s is spread widely 
in the air depth, and so the air density there has a wide distribution.
The variation of the atmospheric density profile changes the
distribution a little, and works mildly on both neutrino and muon productions 
at higher momenta.
With the variation of $\varepsilon=\pm 5~\%$, 
the neutrino flux varies
$\pm$~1.5\% at 1~GeV, $\pm$~1.8\% at 10~GeV, and $\pm$ 2.2\% at 100~GeV/c.
From this variation, we estimate that the error due to the uncertainty of 
atmospheric density profile would be similar to these values or smaller.

The variation of muon flux at lower momenta could be explained by 
the change of production height with the muon decay.
The production height approximated by the constant depth 
($\sim$~100~g/cm$^2$) moves to higher altitude for $\varepsilon > 0$
and lower for $\varepsilon < 0$, and so
the muon flux increases for  $\varepsilon < 0$, 
and decreases for $\varepsilon > 0$ at lower momenta.
The variation with $\varepsilon=\pm$ 5\%, is $\sim \mp 5$\%
at 1~GeV/c and larger at much lower momenta.

Note, there are short-term variations of atmospheric density profile 
due to the change of climate corresponding to $\varepsilon = \pm$ 5~\% 
or more.
They are crucial in the precise comparison of the calculation and observation
of the atmospheric muon flux.
In the following studies,
we calculate the muon fluxes using the observed atmospheric density profile 
for the observation duration, when available.

At the higher momenta ($\gtrsim$~100~GeV/c),
the variation of muon flux by the change of atmospheric density 
profile is smaller than the observation error of muon flux, 
even in the precision measurements.
We use the US-standard '76 atmosphere model in the calculation
for these momenta.

\begin{figure}[tbh]
\includegraphics[width=7cm]{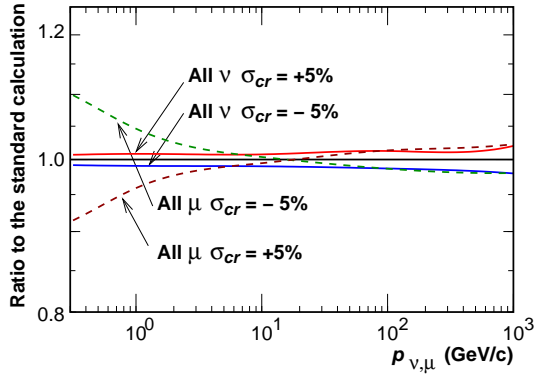}%
\caption{\label{fig:vari-crs}
The variation of atmospheric muon and neutrino fluxes 
resulting from a change of hadronic interaction cross section of
$\sigma\pm 5~\%$.}
\end{figure}

A change in the cross section also changes the first interaction height,
therefore,  we expect a similar variation of 
lepton fluxes to the change in atmospheric density profile.
However, the variation is a little different from that in the
change of atmospheric density profile.
In the static atmosphere model (Eq.~\ref{eq:static-air}),
the mean free path of mesons at the first interaction depth 
is calculated as 
\begin{equation}
\lambda_{m} 
=  \frac{A_{air}}{\sigma_{m}\cdot \rho\cdot N_a} 
= \frac{\sigma_{cr}}{\sigma_m} \cdot h_s
\end{equation}
in real length,
where $\sigma_m$ is $m$-meson interaction cross section.
The production depth of leptons are well approximated 
by the mean free path of cosmic ray in the column density 
calculated as $\Lambda_{cr} =A_{air}/\sigma_{cr} N_a$.
If we assume ${\sigma_{cr}}/{\sigma_m} \sim$ constant,
the change of interaction cross section does not affect the
competition of decays and interactions (Eq.~\ref{eq:int-decay-valence}).
We expect very small variation of atmospheric neutrino flux
with  ${\sigma_{cr}}/{\sigma_m} \sim$ constant in the calculation 
with US-standard '76 or with MSISE90.

We study the effect of the uncertainties of the interaction 
cross sections ratio between cosmic rays and mesons,
on the atmospheric muons and neutrinos.  
In Fig.~\ref{fig:vari-crs} we show the ratios of the 
fluxes calculated with the variation of $\Delta \sigma_{cr}=\pm$~5\%
to the flux calculated with the standard cross sections, 
keeping the ${\sigma_m}$ unchanged.
The interaction cross section of nucleons are varied with 
that of cosmic rays.
As expected, the variation of lepton flux is qualitatively the same with 
the variation due to the change of atmospheric density profile shown 
in Fig.~\ref{fig:vari-by-air}, but quantitatively, the variation
is smaller than that above 1~GeV/c.
The variation of the atmospheric neutrino flux by the change of 
$\sigma_{cr}$ of $\pm 5$~\% is about $\pm$~2~\% at higher momenta,
and is smaller at lower momenta.
The variation of the atmospheric muon flux at the change of
$\sigma_{cr}$ is the same at higher momenta, but is larger 
at lower momenta and in the opposite
direction from that at higher momenta.

\section{\label{sec:comparison}Comparison of calculated and observed muon fluxes}

\begin{figure}[tbh]
\includegraphics[width=7.3cm]{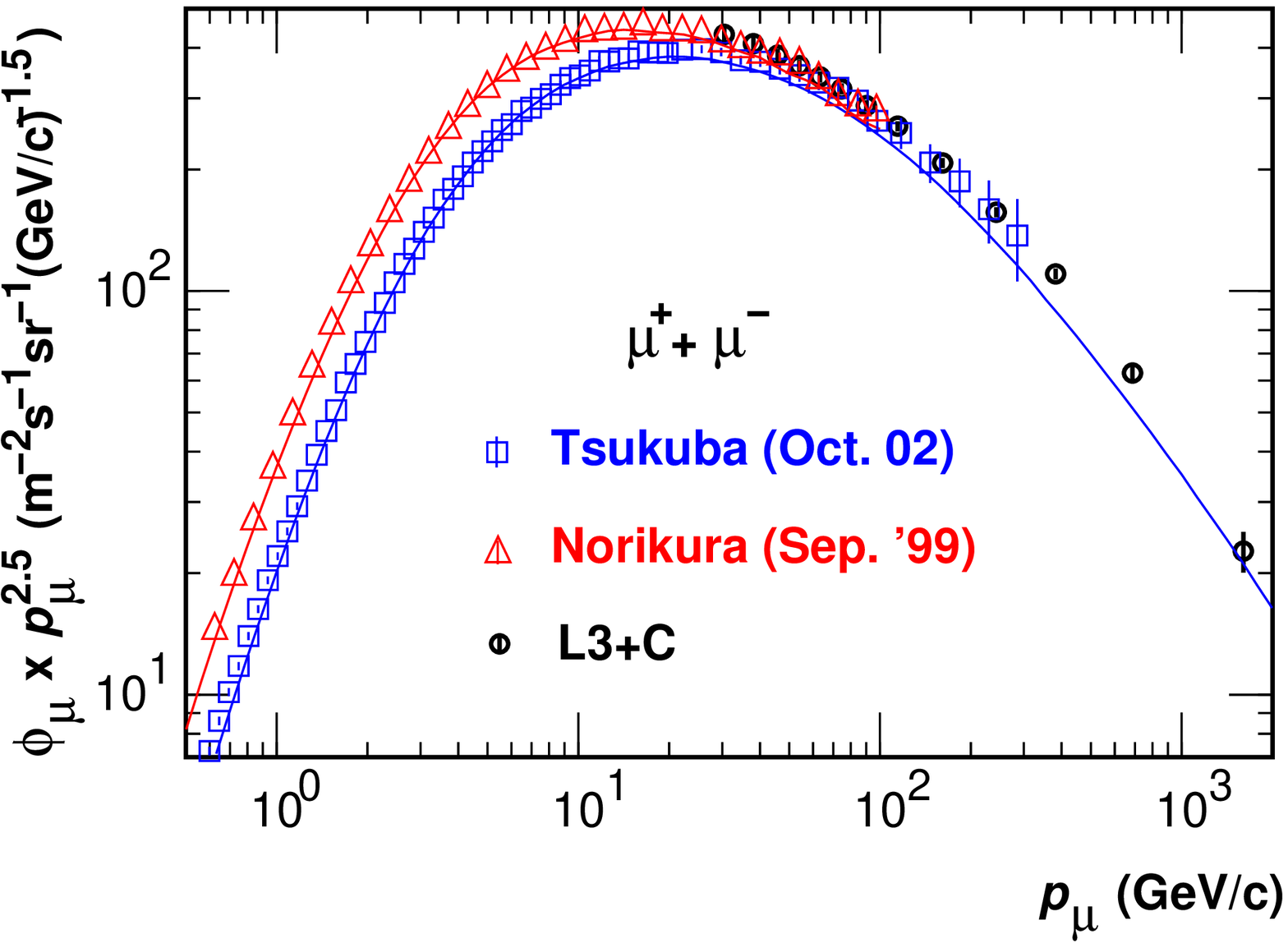}%
~
\includegraphics[width=7cm]{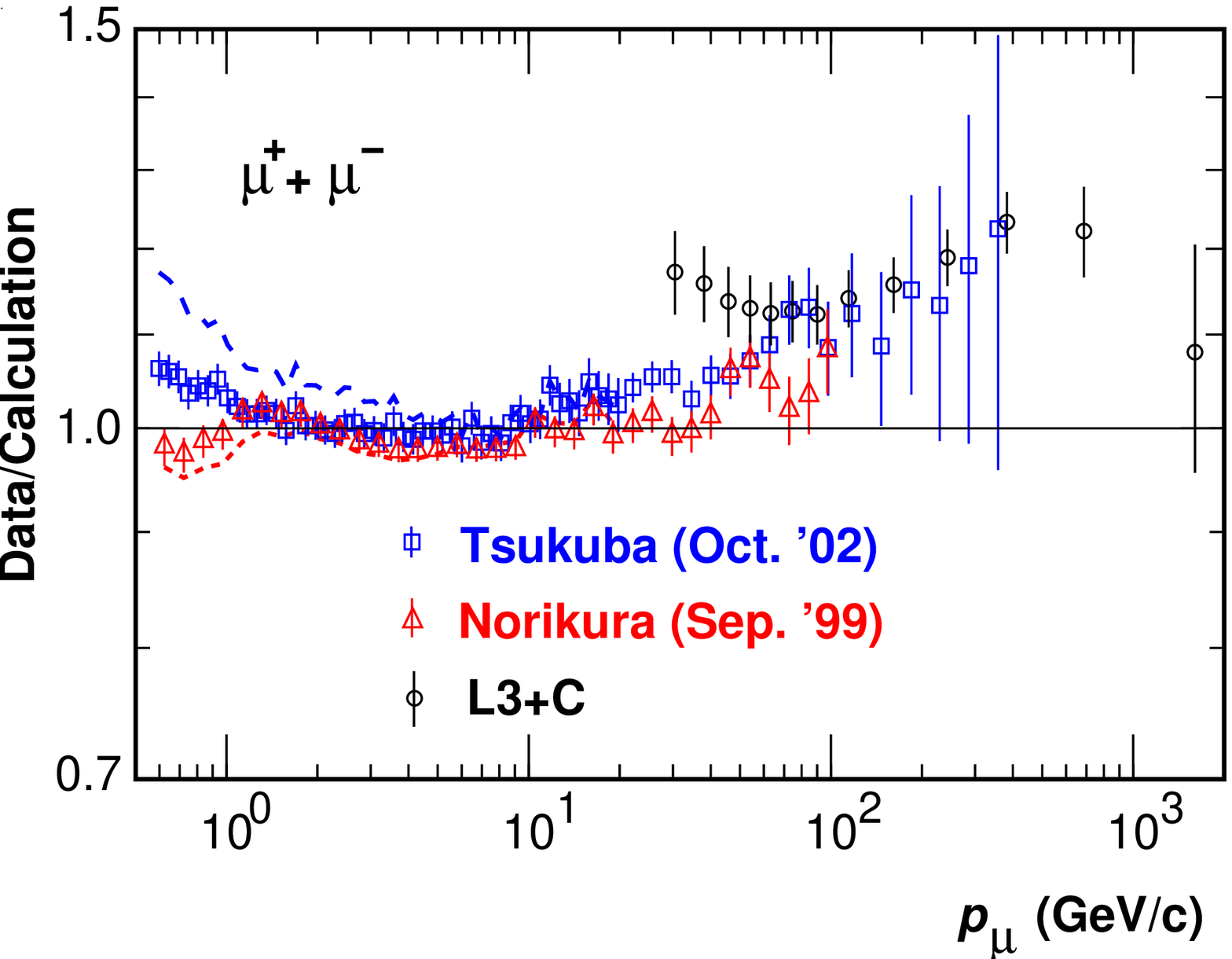}%
\caption{\label{fig:musum04}
Left panel: the muon fluxes measured accurately at 
Tsukuba (Sept.\ 2002)~\cite{BESSTeVpHemu} and 
on Mt.\ Norikura (Oct.\ 1999)~\cite{BESSnorimu} by the BESS group, 
and at CERN by the L3-collaboration (L3+C)~\cite{l3+c}
with the calculated HKKM04 muon fluxes  
for Tsukuba and Mt.\ Norikura.
Right panel: the ratio of the muon flux data to the calculations.
The dashed and dotted lines in the right panel are the same ratios
but calculated with the US-standard '76 atmospheric model for 
Tsukuba and Mt.\ Norikura respectively.
}
\end{figure}

In the left panel of Fig.~\ref{fig:musum04}, we plot 
the muon fluxes measured accurately at 
Tsukuba (Sept.\ 2002)~\cite{BESSTeVpHemu} and 
on Mt.\ Norikura (Oct.\ 1999)~\cite{BESSnorimu} by the BESS group, 
and at CERN by the L3-collaboration (L3+C)~\cite{l3+c}
with the calculated muon fluxes in the HKKM04 scheme for Tsukuba 
and Mt.\ Norikura.
(Details of these experiments are given in Sec.~\ref{sec:obs-mu}.)
In the right panel of Fig.~\ref{fig:musum04},
we plot the ratio $[observation]/[calculation]$ for those precision
measurements to compare the calculation and observation in more detail.
In the calculation of the muon fluxes at Tsukuba and Mt.\ Norikura,
we used the proton and helium fluxes measured by the BESS group 
in the preceding flight carried out within 2 months 
to take into account the solar modulation of cosmic rays correctly.
Also, we used the atmospheric density profile observed 
by the Japan Meteorological Agency~\cite{koso-kishou}
during the experimental periods for them.
The calculations agree with the observations well ($\lesssim$~5\%) 
in the range 1$\sim$30~GeV/c.
Note, the calculation with the US-standard '76 atmospheric
model is also compared with the observed muon data in the figure.

Below 1~GeV/c,
there is a discrepancy between the two $[observation]/[calculation]$ ratios 
calculated for Tsukuba and Mt. Norikura.
This discrepancy might be explained by a different configuration 
of the BESS detector used for those two observations. 
As explained in Sec.~\ref{sec:obs-mu},
an electron/positron component can be distinguished from muons 
in the Norikura observation with the electromagnetic shower counter,
but not in the Tsukuba observation.
Our Monte Carlo study for the observation predicts 
the electron and positron production at the roof of the experimental 
hall for Tsukuba experiment, and it explain at the difference of
calculation and observation at least qualitatively.

In the muon observation at Norikura, 
surviving protons may affect the resultant muon flux.
The $[observation]/[calculation]$ ratio for Norikura shows some structure
between 1 and 3~GeV/c, 
and is systematically smaller than that for Tsukuba above 3~GeV/c.
This might be explained by the treatment of proton contamination
in the positive muon candidates \cite{BESSnorimu}.
At sea level, the proton flux is much smaller than at mountain altitude,
due to the attenuation in the atmosphere between the two altitudes.
Therefore the correction is not necessary for Tsukuba experiment.

Thus, the muon fluxes in 1$\sim$30 GeV/c are well understood by the
HKKM04 calculation with the observed atmospheric density profile.
We may conclude that DPMJET-III can be used to calculate the
atmospheric neutrino flux in 1$\sim$10~GeV region from the conclusion
in Sec.~\ref{sec:mn-relation}.
However,
the muon flux calculated in the HKKM04 scheme
is clearly smaller than those observed by the precision measurements
above 30 GeV/c.
At those momenta, 
it is difficult to understand the difference with the uncertainty of the
physical inputs, such as the atmospheric density profile,
other than the primary cosmic ray model or the hadronic interaction model
above 100~GeV.
We note similar deficit is observed in the comparison of calculation
with DPMJET-III and the observation of atmospheric gamma ray 
flux~\cite{atm-gamma}.

\begin{figure}[tbh]
\includegraphics[width=7cm]{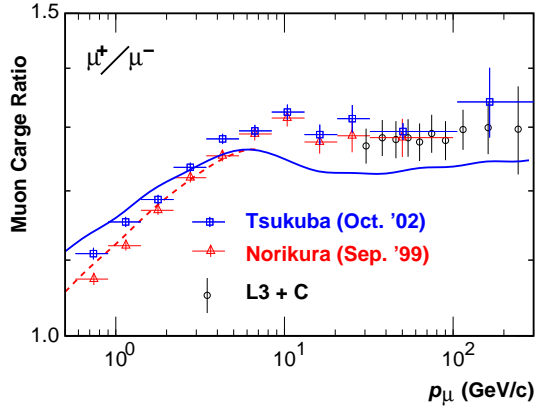}
\caption{
\label{fig:mu-charge-ratio0}
 Comparison of calculation and observation for muon charge ratio.
Upper data and the solid line show the muon charge ratio at 
ground level (Tsukuba and L3+C).
Lower data and the dashed line show the same quantities for 
Mt.\ Norikura (2770m a.s.l.).
}
\end{figure}

In Fig.~\ref{fig:mu-charge-ratio0}, we show the comparison of the
calculated muon charge ratio with the observed ones.
We find the agreement of the calculation and observation 
are better than 10~\% in the all energy region.
However, the muon charge ratio is much more robust observation quantity,
and reflects almost directly the $\pi^+/\pi^-$ ratio of the 
hadronic interaction model.
The difference indicates an error in the 
charge ratio of the $\pi$ production in DPMJET-III.

The difference of muon charge ratio at different observation levels
comes from the muon energy loss in the atmosphere.
Most of muons are produced at higher altitudes than Norikura,
and they are observed as slightly higher momenta muons
at Norikura altitudes than at sea level, due to the muon energy loss.

\section{\label{sec:modification} Modification of DPMJET-III}

Here we consider the modification of DPMJET-III~\cite{Roesler:2000he}, 
without discussing the dynamics of the hadronic interaction,
and actually apply the modification to the 
``inclusive DPMJET-III''~\cite{HKKM2004}.
The inclusive DPMJET-III is constructed from the output of the
original DPMJET-III, so that it reproduces the secondary
spectra of the original DPMJET-III in an inclusive way.
In this interaction model, the conservation laws are
violated in each interaction, but are
satisfied in the statistical way.
Therefore, it is not useful to reproduce an event
caused by a single cosmic ray, but is much faster than 
the original interaction model. The computation speed is
very important in the calculation of the atmospheric neutrino 
flux.

Note, the region of $x \gtrsim 0.1$ in the  $\pi$ and $K$
production is the most responsible for the atmospheric muons and neutrino 
fluxes (see Sec.\ref{sec:mn-relation}).
We modify the gradient of the secondary spectra 
to cause changes at $x \sim 0.1$ for $\pi$'s and $K$'s 
without touching the multiplicities.
Therefore, the quantum numbers are conserved automatically.
For the magnitude of modification, 
we use the ratio of the average energy before and after the modification.

We assign a modification parameter to a valence quark of the projectile, 
and consider the same magnitude of modification for the secondary particles 
which have the same valence quark as the projectile.
In $p + Air$ interactions, the change of average energy are assigned as:
\begin{equation}
\label{eq:pAir-param}
\begin{array}{llllll}
<E_{\pi^+}> &=& (1 + c_u) <E^0_{\pi^+}>&&&(u\bar d)\\
<E_{\pi^-}> &=& (1 + c_d) <E^0_{\pi^-}>&&&(d \bar u)\\
<E_{\pi^0}> &=&(1 + (c_u + c_d)/2) <E^0_{\pi^0}>
         &&\hspace{1cm}&((u\bar u - d\bar d)/2)\\
<E_{K^+}> &=& (1 + c_u) <E^0_{K^+}>&&&(u \bar s)\\
<E_{K^-}> &=&  <E^0_{K^-}>         &&&(s \bar d)\\
<E_{K^0}> &=& (1 + c_d) <E^0_{K^0}>&&&(d \bar s)\\
<E_{\bar K^0}> &=& <E^0_{\bar K^0}>&&&(s \bar d)\\
\end{array}
\end{equation}
Here, $c_u$ and $c_d$ are the modification parameters assigned 
to the $u$ and $d$ quarks respectively, 
and the $<E^0_i>$ is the average energy in the original DPMJET-III
for the $i$ particle.
As the $K^0$ and $\bar K^0$ oscillate quickly, 
their average energies are effectively modified as 
$<E_{K^0, \bar K^0}> = (1 + c_d/2) <E^0_{K^0}>$.
Note, 
the modification of the nucleon spectra is determined after
the modification for mesons are determined, so that the total energy
is conserved to be equal to that of the projectile.
These assumptions and parameterization naturally relate the 
$K$ and $\pi$ productions through the parameters assigned 
for the $u$ and $d$ quarks.

For the $n + Air$ interactions, we assume iso-symmetry, 
or that the parameter for the $d$-quark in $n+Air$ interactions 
is equal to the $c_u$ in Eq.~\ref{eq:pAir-param},
and that for the $u$-quark is equal to the $c_d$.
As $p+Air$ and $n+Air$ are the major interactions in 
the cosmic ray propagation process in air, 
$c_u$ and $c_d$ are the two major parameters (Fig.~\ref{fig:n2n}).

For the energy dependence of $c_i$'s,
we consider polyline functions with kinks at
1, 3.16, 10, 31.6, 100, $\ldots$ GeV.
However, there are only a small number of data points above 100~GeV/c,
and the uncertainty of the primary flux data is large there.
We simply assume:
\begin{equation}
\label{eq:mod-par}
c_{i} = a_{i} \cdot \log_{10}({E_{proj} /10~{\rm GeV}})
\end{equation}
above 316 ~GeV for $i=u$ and $d$.
Then, we tuned the ${c_u}$'s and ${c_d}$'s at the kink points and 
$a_u$, $a_d$ 
to minimize the difference between calculations and observations.
In this study,
we used the muon flux data from L3+C at CERN above 60~GeV/c,
and BESS at Tsukuba for all the momentum region. 
Note, 
the BESS data did not suffer from the effects of overlying material.

\begin{figure}[tbh]
\includegraphics[width=7cm]{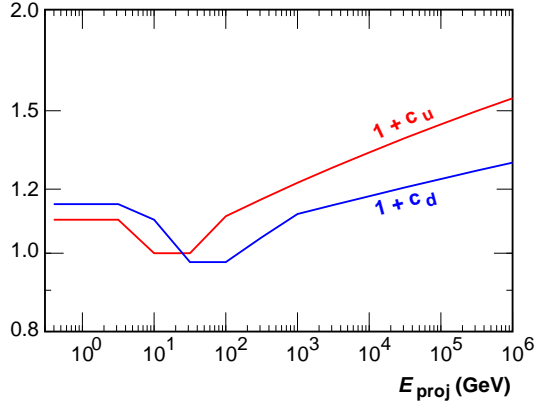}%
\caption{\label{fig:mod-par}
The best modification parameters, $c_u$'s and $c_d$'s, 
as a function of $E_{proj}$.
}
\end{figure}
\begin{figure}[tbh]
\centerline{
\includegraphics[width=7cm]{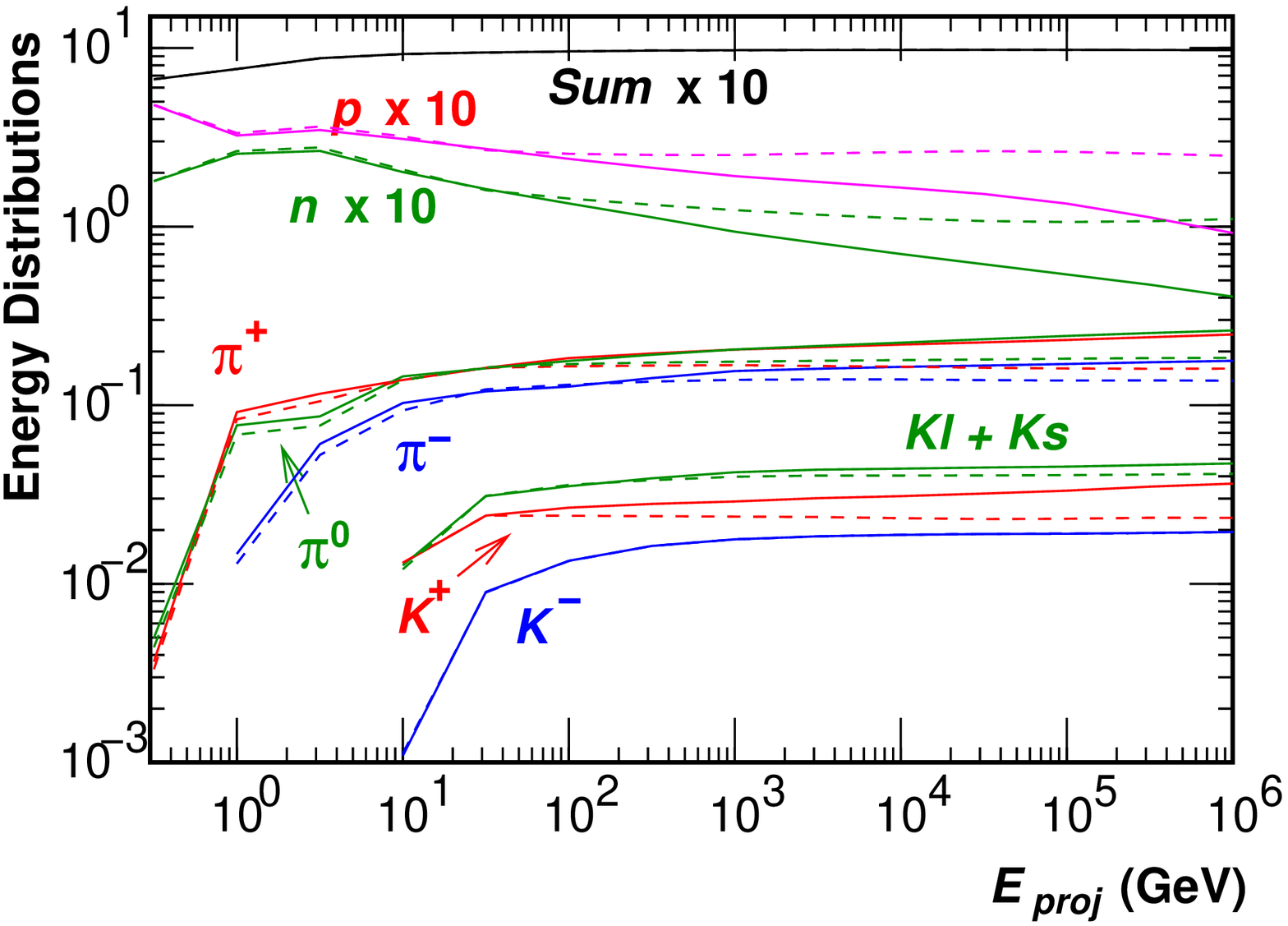}%
\hspace{5mm}
\includegraphics[width=7cm]{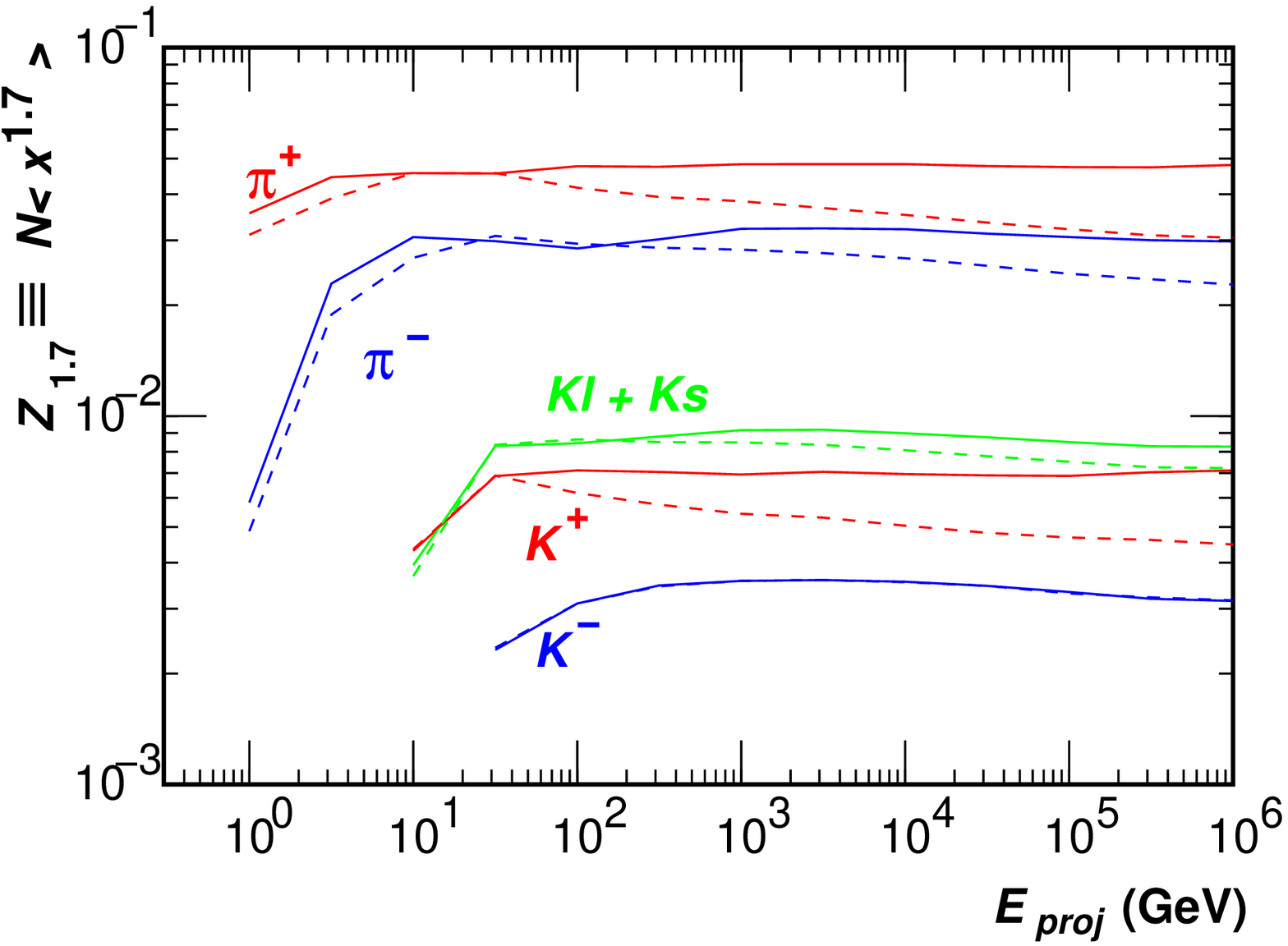}%
}
\caption{\label{fig:e-distribution}
Left panel: The ratio of the average energy of secondary particles 
to the projectile energy.
Right panel: the $Z$-factors for different kinds of particles.
The $Z$-factor is defined in the text.
In both panels, the solid lines show the modified ones,
and dashed lines show the original ones.
}
\end{figure}

With these procedures, we find the $c_u$'s and $c_d$'s 
connected by the polylines in Fig.~\ref{fig:mod-par} give the best result.
The kinks seen at around 10 GeV are 
due to the connection to the NUCRIN interaction model in 
Figs.~\ref{fig:mod-par} and \ref{fig:e-distribution}.
Note, the differences of $c_u$'s and $c_d$'s result from the 
difference in the muon charge ratio in the observation and 
calculation (Fig. \ref{fig:mu-charge-ratio0}).
We call thus modified interaction model as ``modified DPMJET-III''.

We compared 
the energy distributions to the secondary particles of the
original and modified DPMJET-III
in the left panel of Fig.~\ref{fig:e-distribution},
and $Z$-factors in the right panel.
The $Z$-factor is defined as
\begin{equation}
Z_{i} \equiv  N_i <x_{i}^{1.7}> ,
\hspace{5mm} {\rm and}\hspace{5mm} x_{i} \equiv \frac {p_i}{p_{proj}} ,
\label{eq:z17}
\end{equation}
where $N_i$ is the multiplicity and $x_{i}$ is the scaling variable 
used in Sec.~\ref{sec:mn-relation}
for the $i$ secondary particle $(\pi^+, \pi^-, ... )$.
The power 1.7 is approximately equal to the 
integral spectrum index of the cosmic ray protons (1.71 in the 
primary flux model used by HKKM04). 
The $Z$-factor plays an essential role in the analytic calculation of
the atmospheric muon and neutrino flux at higher energies.
We find the $Z$-factors in modified DPMJET-III have flatter energy 
dependences than the original one above 100~GeV, 
as is suggested by the scaling hypothesis.

\begin{figure}[tbh]
\centerline{
\includegraphics[width=5cm]{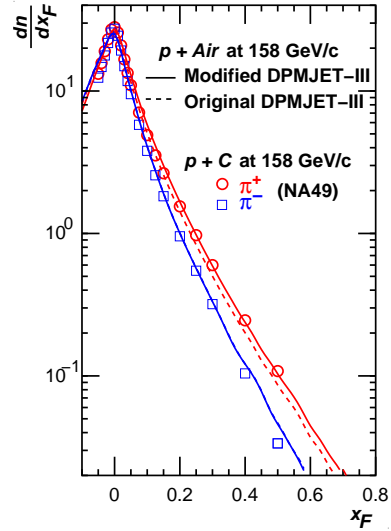}%
}
\caption{\label{fig:na49}
The comparison of the $\pi$'s production spectra of the original 
and modified DPMJET-III in $p+Air$ interaction with that in $p+C$ from 
the NA49 experiment at 158~GeV/c. 
Note, 
the production spectrum of $\pi^-$ of the modified DPMJET-III 
at 158~GeV/c is almost the same as that of the original DPMJET-III.
}
\end{figure}

In Fig.~\ref{fig:na49}, we compared the $x_F$ spectra of 
$\pi$'s production of $p+Air$ interactions in the original and 
modified DPMJET-III with that of $p+C$ interactions in
the NA49 experiment at 158~GeV/c \cite{na49-data}.
In this comparison, we use the
Feynman scaling variable defined as $x_F = p_\parallel/2\sqrt{s}$
in the CM-frame of a projectile and a nucleon in the target nucleus.
Note, the scaling variable $x$ we used in Sec.~\ref{sec:mn-relation}
is defined a little differently from the $x_F$ using the momenta 
at the rest frame of Air nuclei.
However, both definitions are almost equivalent for $x_F > 0.1$
at 158~GeV/c.
We find the modified DPMJET-III reproduce the production 
spectra at $x_F \gtrsim$~0.2 better than the original DPMJET-III.

\section{\label{sec:modified-calculation}The calculations with the modified interaction model}

\begin{figure}[tbh]
\includegraphics[width=7cm]{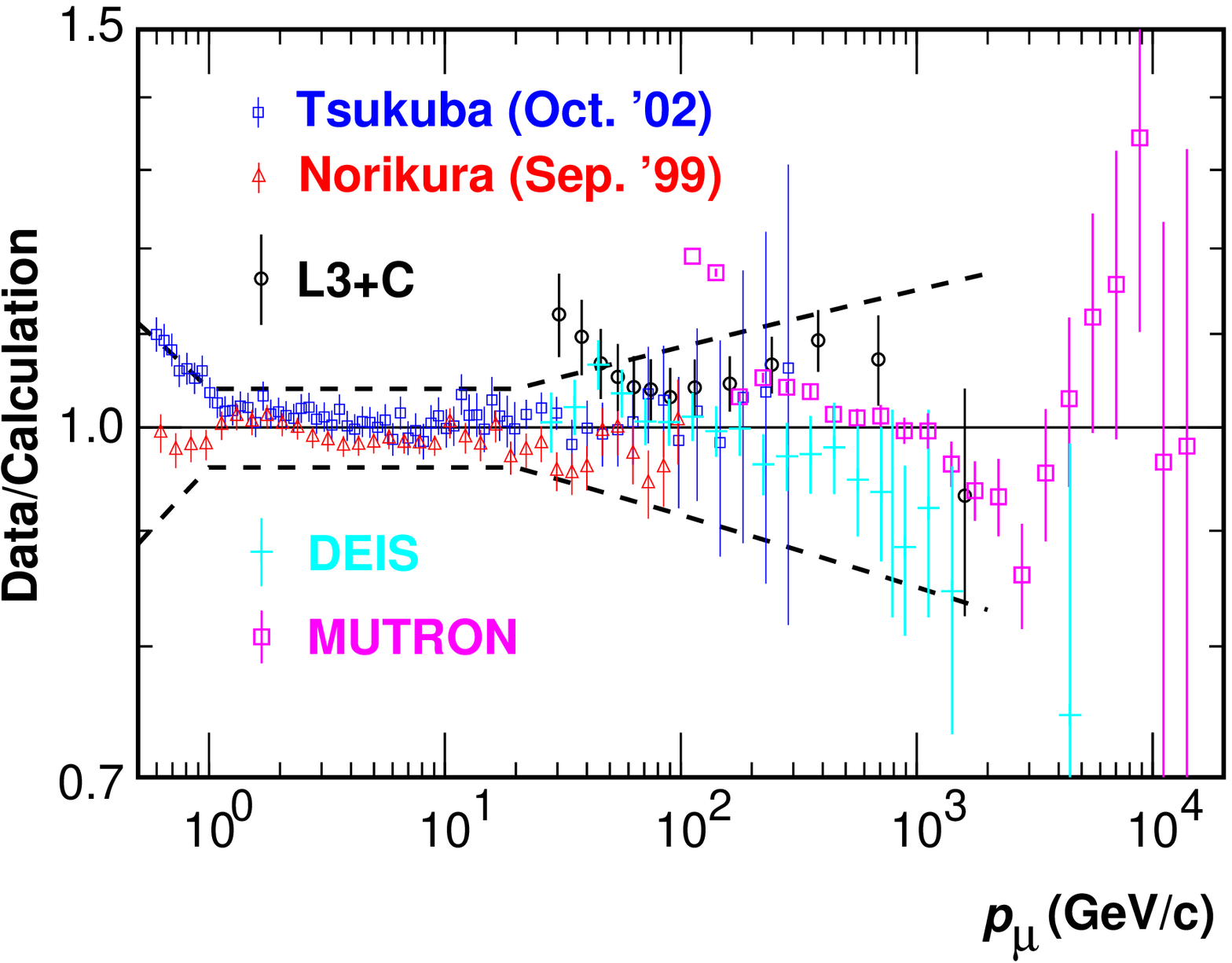}%
\hspace{5mm}
\includegraphics[width=7cm]{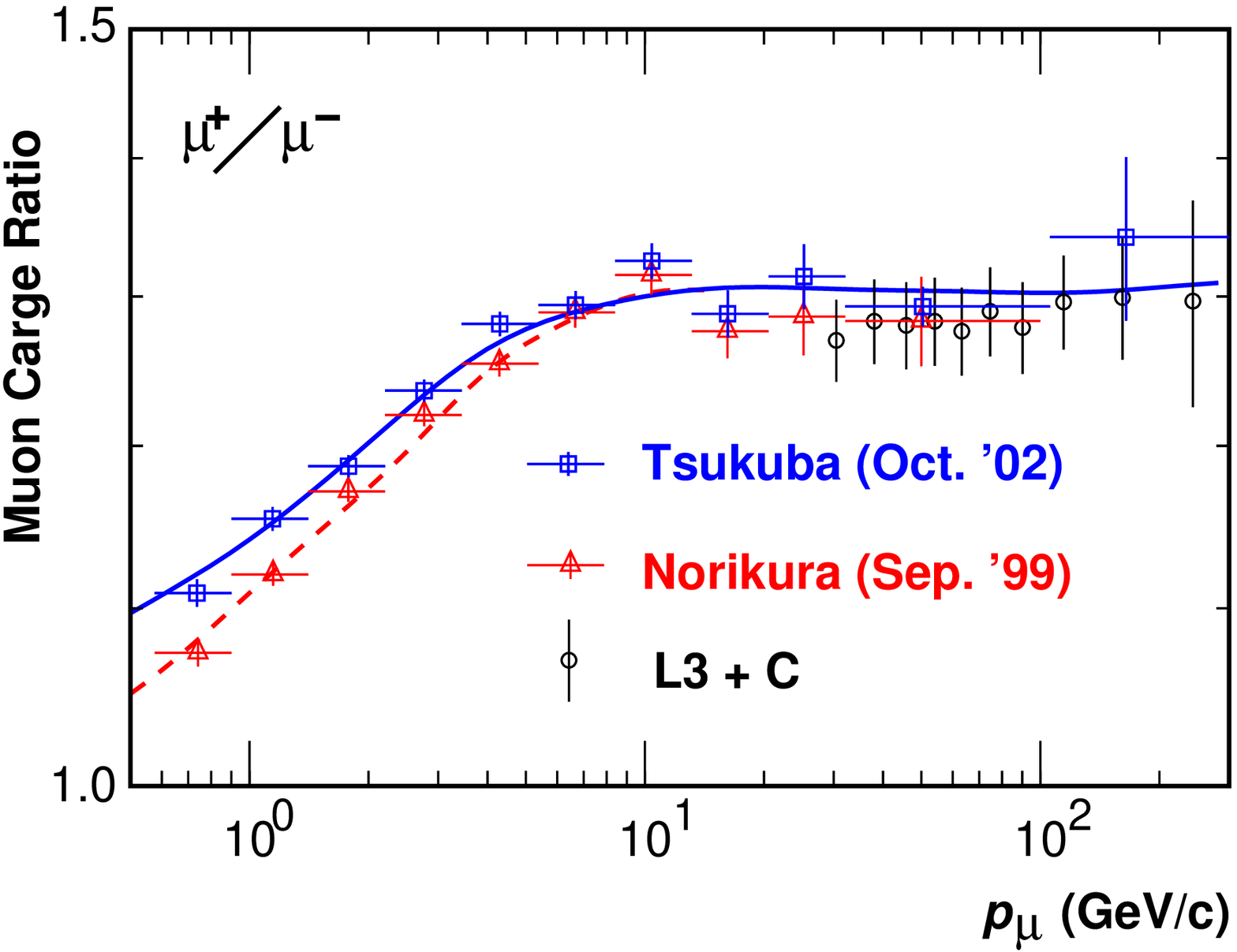}
\caption{\label{fig:modified-mu}
The comparison of calculated muon fluxes with the modified interaction
model and the observed ones.
The dashed line is the sum of the errors in data and residuals by the modification (Eq.~\ref{eq:errmu}).
Left panel: for the muon flux sum ($\mu^+ + \mu^-$), and 
Right panel: for charge ratio ($\mu^+ / \mu^-$).
Note, the systematic errors for DEIS and MUTRON are not included 
in the error bars.
}
\end{figure}

In this section, we calculate the muon fluxes for the observations at 
Mt.\ Norikura, Tsukuba and CERN with the modified DPMJET-III
described in the previous section. 
Except for the interaction model,
we exactly repeat the calculations in Sec.~\ref{sec:comparison},
i.e. 
we use the primary flux model based on the measurement within 2 months, 
and observed atmospheric density profile during
the experimental period for Mt.\ Norikura and Tsukuba.
The muon flux sum ($\mu^+ + \mu^-$) is compared in ratio 
$[observation]/[calculation]$ in the left panel of Fig.~\ref{fig:modified-mu},
and the muon charge ratio in the right panel. 
In the left panel,
we also plotted the ratio for the horizontal muon flux data 
observed by the DEIS~\cite{Allkofer:1985ey} and MUTRON~\cite{Matsuno:1984kq}
experiments.

Comparing Fig.~\ref{fig:modified-mu} with Figs.~\ref{fig:musum04} 
and \ref{fig:mu-charge-ratio0},
we find the agreement of calculation and observation of the muon fluxes is 
greatly improved by the modification in 30$\sim$300~GeV/c for 
muon fluxes at Mt.\ Norikura and Tsukuba
and between 60~GeV$\sim$2~TeV for the CERN experiment.
We summarized the remaining differences between calculations 
with the modified DPMJET-III and 
observations, including the experimental errors, as,
\begin{equation}
\label{eq:errmu}
\delta\Phi_\mu = \Bigg\{
\begin{array}{l c}
0.04-0.24\cdot \log(\frac{1~\rm GeV/c}{p_\mu}) &~~p_\mu < 1~{\rm GeV/c,}\\
0.04 &~~1~{\rm GeV/c}< p_\mu < 20~{\rm GeV/c, ~and}\\
0.04+0.065\cdot \log(\frac{p_\mu}{20~\rm GeV/c})&~~p_\mu > 20~{\rm GeV/c}~,\\
\end{array}
\end{equation}
and plotted these in Fig.~\ref{fig:modified-mu} with dashed lines.
We also find DEIS and MUTRON data agree with the calculation
in the momentum ranges of 60$\sim$600~GeV/c and 200~GeV$\sim$2~TeV/c 
respectively, and are well inside the dashed lines.
Note, the systematic error for DEIS and MUTRON are not included 
in the error bars.
The modified DPMJET-III should be able to calculate the atmospheric neutrino 
flux, at least for the $\pi$-decay, with good accuracy above 1~GeV,
from the study described in Sec.~\ref{sec:mn-relation}.

\section{\label{sec:summary}Summary}

In this paper,
we have studied the hadronic interaction for the calculation of
atmospheric neutrino flux,
using atmospheric muon flux data observed by 
precision measurements.

We summarized the muon data from the precision measurements,
and selected the data from BESS and L3+C for our study.
There are other potentially useful data,
such as muon observations at balloon altitudes by BESS,
or for horizontal directions by MUTRON or DEIS.
However, the former still suffers from small statistics,
and the latter do not clearly quantify the systematic errors in 
their reports.

Then, we studied the $\pi$ and $K$ productions in the
hadronic interactions of cosmic rays and air nuclei relevant to 
the atmospheric muons and neutrinos.
In this study we manipulated analytic expressions, 
but the actual calculations were carried out in the Monte Carlo 
simulation using the HKKM04 calculation code.
With the Monte Carlo data 
being interpreted with the use of the analytic expressions,
we found the atmospheric muon and neutrino fluxes
originated from the $\pi$ decay have the relation,
$\Delta\Phi_\mu/\Phi_\mu \simeq \Delta\Phi_{\nu_\mu}/\Phi_{\nu_\mu} 
\simeq \Delta\Phi_{\nu_e}/\Phi_{\nu_e}$ for $\gtrsim$~1 GeV.
This relation is useful to study the error in the hadronic
interaction model using the atmospheric muon data.

As original DPMJET-III can reproduce the muon flux data
in 1$\sim$30~GeV/c with the HKKM04 calculation scheme,
we may say that DPMJET-III is good interaction model to 
calculation the atmospheric neutrino flux in the 1$\sim$10~GeV range.
Note that $\pi$'s are the main source both for atmospheric muons and
neutrinos in this energy region.
However, the observed muon data show a sizable deviation  
from the calculated muon flux above 30~GeV/c
and in the muon charge ratio,
suggesting room for improvement in the DPMJET-III model.
We note similar deficit is observed in the comparison of calculation
with DPMJET-III and the observation of atmospheric gamma ray 
flux~\cite{atm-gamma}.

We tried an improvement of the interaction model based on 
the quark parton model, and 
applied the modification to the ``inclusive DPMJET-III''.
We have tuned the secondary spectra of the hadronic interactions,
so that the calculation reproduces the observed muon fluxes accurately.
As a result,
the muon fluxes calculated with the modified interaction model 
agree very well with the observed muon flux data,
up to $\sim$1~TeV for both vertical and horizontal directions in the
flux sum ($\mu^+ + \mu^-$) and up to 300~GeV/c in the charge ratio.
The muon flux data for horizontal directions are potentially 
useful to examine the propagation code of cosmic rays in the 
atmosphere.
However, the systematic error of the available muon flux data 
for horizontal directions ware not studied well.
We just show the comparison of calculation in the figure in this 
paper.

The calculation of atmospheric neutrino flux, and 
the robustness of our modification to the DPMJET-III model,
will be described in the following paper~\cite{hkkms2006}.
Note, the $K$ productions in the hadronic interactions
are naturally modified through the modification for $u$ and $d$-quarks.
However, the modification of $K$ productions weakly couple to 
the muon fluxes, and are
not tested in the comparison of calculation and observation of muon 
flux below $\sim$1~TeV.
The uncertainty of $K$ productions and the results in atmospheric 
neutrino flux calculation is also discussed in the following paper.

\section{Acknowledgments}
We greatly appreciate the contributions of 
J.~Nishimura and A.~Okada to this paper. 
We are also grateful to R.~Engel, P.~Lipari, K.~Abe, S.~Haino 
and P.G.~Edwards for useful discussions and comments. 
We thank the ICRR of the University of Tokyo, for support.
This study was supported by Grants-in-Aid, Kakenhi (12047206), from
the Ministry of Education, Culture, Sport, Science and Technology 
(MEXT) in Japan.

\bibliography{mnflux1}
\end{document}